\documentclass[twocolumn]{aastex631}
\usepackage{amsmath}
\usepackage{amsthm}
\usepackage{amssymb}
\usepackage{graphicx}
\usepackage{physics}
\usepackage{color}
\usepackage{soul}
\usepackage{hyperref}
\usepackage{bm}
\usepackage{footmisc}

\newcommand{\angstrom}{\mbox{\AA}}
\usepackage{booktabs}
\usepackage{overpic}

\shorttitle{Star Formation Efficiency During Reionization}
\shortauthors{Sipple \& Lidz}

\newcommand{\dif}{\text{d}}

\begin{document}

\title{The Star Formation Efficiency During Reionization as Inferred from the Hubble Frontier Fields}

\author{Jackson Sipple}
\affiliation{Center for Particle Cosmology, Department of Physics and Astronomy, University of Pennsylvania, Philadelphia, PA 19104, USA}

\author[0000-0002-3950-9598]{Adam Lidz}
\affiliation{Center for Particle Cosmology, Department of Physics and Astronomy, University of Pennsylvania, Philadelphia, PA 19104, USA}

\correspondingauthor{Jackson Sipple}
\email{jsipple@sas.upenn.edu}

\begin{abstract}
A recent ultraviolet luminosity function (UVLF) analysis in the Hubble Frontier Fields, behind foreground lensing clusters, has helped solidify estimates of the faint-end of the $z \sim 5-9$ UVLF at up to five magnitudes fainter than in the field.
These measurements provide valuable information regarding the role of low luminosity galaxies in reionizing the universe and can help in calibrating expectations for JWST observations. We fit a semi-empirical model to the lensed and previous UVLF data from Hubble. This fit constrains the average star formation efficiency (SFE) during reionization, with the lensed UVLF measurements probing halo mass scales as small as $M \sim 2 \times 10^9 {\rm M}_\odot$. The implied trend of SFE with halo mass is broadly consistent with an extrapolation from previous inferences at $M \gtrsim 10^{10} {\rm M}_\odot$, although the joint data prefer a shallower SFE. This preference, however, is partly subject to systematic uncertainties in the lensed measurements. 
Near $z \sim 6$ we find that the SFE peaks at $\sim 20 \%$ between $\sim 10^{11}-10^{12} {\rm M}_\odot$. Our best fit model is consistent with Planck 2018 determinations of the electron scattering optical depth, and most current reionization history measurements, provided the escape fraction of ionizing photons is $f_{\rm esc} \sim 10-20\%$. The joint UVLF accounts for nearly $80\%$ of the ionizing photon budget at $z \sim 8$. Finally, we show that recent JWST UVLF estimates at $z \gtrsim 11$ require strong departures from the redshift evolution suggested by the Hubble data.
\end{abstract}

\setlength{\footnotemargin}{\parindent}

\section{Introduction}
\label{sec:intro}

Measurements of the galaxy luminosity function provide a fundamental input for models of galaxy formation and evolution, with Hubble Space Telescope (HST) observations probing the ultraviolet luminosity function (UVLF) out to $z \sim 10$ and ongoing JWST studies reaching still earlier phases
in our cosmic history. In current models of structure formation, dark matter halos collapse under gravity and galaxies form as gas falls into these halos, cools, and fragments to form stars \citep{White78}. This process is thought to be regulated by feedback from supernova explosions, stellar radiation, and via the energy injected into the gas from active galactic nuclei (AGN) outflows and radiation. While the abundance of dark matter halos as a function of mass and redshift is relatively well understood from N-body simulations and analytic theory \citep{Sheth:2001dp}, galaxy formation and the feedback processes which regulate it, are challenging to model. 

The UVLF measurements provide important empirical guidance for efforts to model galaxy formation: after anchoring to the well understood abundance of dark matter halos (the ``halo mass function''), the UVLF can be used to determine correlations between the UV luminosity of a galaxy and the mass of the dark matter halo which hosts it. This, in turn, is closely connected to the star formation efficiency (SFE), which
quantifies the fraction of the baryons accreting onto a halo that are converted into stars. In so-called ``semi-empirical models'', these
connections between UV luminosity, star formation rate, and host halo mass are extracted from UVLF observations and can be used 
to cross-check or constrain galaxy formation models (e.g. \citealt{Vale:2004yt,Behroozi:2014tna,Sun2016}). 

Among other topics, the inferred star formation efficiencies have important implications for our understanding of the Epoch of Reionization (EoR). The EoR is the time period during which the first luminous sources form, photo-ionize surrounding neutral hydrogen, and gradually fill the universe with ionized gas \citep{Loeb2013}. Important goals for reionization studies include determining the volume-filling factor of ionized gas in the intergalactic medium (IGM) as a function of redshift and characterizing the properties of the ionizing sources. Joint measurements of the reionization history and the galaxy UVLF can address whether we have an accurate census of the sources of reionization, or whether missing populations of faint galaxies, accreting black holes, or other more exotic possibilities are
required. 

Measurements of the UVLF from HST combined with estimates of the electron scattering optical depth to cosmic microwave background (CMB) photons, $\tau_e$, from WMAP \citep{Hinshaw13} and Planck \citep{planck2018} have started to address this question. Additional information comes from measurements of the average ionization fraction as a function of redshift \citep{McGreer:2014qwa,Davies:2018yfp,Mason:2017eqr,Mason:2019ixe,Umeda23}, and inferences of the emissivity of ionizing photons from the Lyman-alpha (Ly-$\alpha$) forest \citep{Becker21,Gaikwad:2023ubo}. A number of previous studies have found that recent HST galaxy UVLF measurements are in accord with Planck $\tau_e$ estimates provided that a significant fraction -- around $\gtrsim 10-20\%$ -- of ionizing photons escape their host galaxies (e.g. \citealt{Robertson:2015uda,Mashian2016,Sun2016,Yung2020}). Since the escape fraction is hard to measure, especially in reionization-era galaxies, it remains unclear whether this requisite escape fraction is reasonable. 
There are also still sizable uncertainties related to how one extrapolates the UVLF down the faint-end and in redshift, as well as in the spectral shape of the UV emission from galaxies. For example, there is not yet a consensus regarding the role of low-luminosity galaxies in reionization, with some works arguing that faint
$(M_{UV} \gtrsim-18$) galaxies contribute minimally \citep{Naidu_2020}, while
other studies suggest that these sources are dominant \citep{Lewis_2020}.

The intensity of UV radiation and its redshift evolution also have important consequences for 21 cm surveys. Specifically, UV photons redshifting into Lyman-series resonances are thought to couple the 21 cm spin temperature to the gas temperature, allowing the 21 cm signal to be observable in absorption against the CMB during the Cosmic Dawn (before e.g. X-rays heat the gas temperature above the CMB temperature leading to an emission signal; \citealt{Furlanetto:2006jb}.)  

The long-anticipated JWST is starting to revolutionize our understanding of early galaxy populations and further address these questions in some detail. Indeed, after only months of operation, early JWST detections of $z \sim 11-17$ photometric candidate galaxies already suggest a surprisingly large abundance of UV-bright galaxies in the early universe (e.g. \citealt{Donnan22,Naidu22,bouwens2023evolution,Harikane_2023}).
Additionally, the stellar mass estimates for some high redshift JWST candidate galaxies appear quite a bit higher than expected \citep{Labbe23,boylankolchin2023stress}. On the other hand, the JWST analyses are still in their early stages -- for one, the total number of candidates identified and the volume surveyed are still small and so statistical uncertainties remain large.

In this paper, we focus our attention on interpreting HST UVLF measurements including a recent analysis of the UVLF in the Hubble Frontier Fields (HFF) from \cite{Bouwens22} (hereafter B22). The HFF program \citep{Coe15,Lotz17} took deep multi-band images around six galaxy clusters (and six parallel fields): the foreground galaxy clusters lens distant background galaxies, which allows measurements of the reionization-era UVLF
at unprecedentedly small luminosities. As we review in \S \ref{sec:uvlf_data}, there are a number of systematic
concerns with lensed UVLF measurements: however, the community has made progress in accounting for such
worries and B22 presents a comprehensive analysis with a large number of reassuring cross-checks. 
The B22 measurements probe absolute UV magnitudes as faint as $M_{\rm UV} = -12.75$, nearly
five magnitudes fainter than probed in UVLF measurements towards typical regions (i.e., without the magnification boost from a foreground cluster). The B22 HFF analysis can also be combined with an earlier analysis from \cite{Bouwens21} (hereafter B21) which determined the UVLF in typical regions. Naturally, the B21 measurements span more cosmic volume, contain a larger sample of galaxies, and hence provide better estimates of the bright end of the UVLF than does B22. Taken together, the combined B21+B22 data provide estimates of the UVLF across a large dynamic range in luminosity and over a decent fraction of the EoR (out to $z \sim 10$). However, the combined B21+B22 results have yet to be compared with semi-empirical UVLF models. 

The goal of the present work is hence to fit semi-empirical models to the joint B21 and B22 UVLF measurements. Of particular interest here is the increased leverage in luminosity from the faint-end UVLF measurements of B22, which allow determinations of the SFE in small mass halos.
We then explore the resulting implications for the reionization history of the universe,
the electron scattering optical depth, measurements of the ionizing emissivity from the Ly-$\alpha$ forest, and for the timing of the 21 cm absorption signal. We will also compare our HST-calibrated models to early results from JWST. 

Specifically, the outline of this paper is as follows. In \S \ref{sec:uvlf_data} we briefly summarize the UVLF data from B21+B22. In \S \ref{sec:Method} we describe our semi-empirical modeling approach and our procedure for fitting models to the UVLF measurements. We present the results of our fits in \S \ref{sec:results}, while \S \ref{sec:implications} explores their implications, including for: the SFE versus halo mass (\S \ref{sec:sfe_results}), the reionization history (\S \ref{sec:ionization_hist}), our census of the sources of reionization (\S \ref{sec:census}), the ionizing emissivity (\S \ref{sec:emissivity}), the electron scattering optical depth (\S \ref{sec:tau_e}), and the timing of the 21 cm absorption signal (\S \ref{sec:wf_coupling}). In \S \ref{sec:jwst} we compare our models with early JWST results. Finally, \S \ref{sec:alt_models} considers extensions of our fiducial model, while \S \ref{sec:conclusion}
presents our conclusions and mentions possible future research directions.
Throughout, we assume a LCDM model described by the following parameters: $\Omega_m=0.31$, $\Omega_\Lambda=0.69$, $\Omega_b=0.049$, $h=0.677$, $\sigma_8=0.818$, $n_s=0.96$, in agreement with Planck 2018 results \citep{planck2018}.  

\section{Ultraviolet Luminosity Function Data}
\label{sec:uvlf_data}

In this work, we primarily compare models with the state-of-the-art pre-JWST UVLF measurements from the HST, using the analyses behind foreground lensing clusters from the HFF in B22 and estimates in the ``field'' (i.e. in typical regions sometimes referred to as ``blank-fields'') from B21. Here we briefly summarize the properties of these data sets, and describe potential systematic concerns with the UVLF estimates behind lensing clusters. 

The HFF program obtained deep multi-band images around six galaxy clusters with well-characterized mass distributions, along with observations in six parallel fields \citep{Coe15,Lotz17}. This program was designed to probe the faint end of the UVLF during the EoR, exploiting gravitational lensing
magnification by foreground galaxy clusters to access distant and low luminosity background galaxies. Although this technique
is powerful, fully robust UVLF measurements behind lensing clusters require mitigating or accounting for a number of
potential sources of systematic error. These include the need: to accurately subtract light contamination from the lensing cluster itself, to properly account
for the size distribution of background galaxies (with the source size impacting the detection efficiency behind a lensing cluster), and to quantify the effect of uncertainties in the lensing models (see e.g. B22 and references therein). A great deal of progress, however, in addressing these concerns has been made through a number of HFF analyses over the years (e.g. \citealt{Atek15,Atek18,Bouwens17b,Bouwens22,Bhatawdekar19,Castellano16,Ishigaki18,Livermore17,Yue2018}). 

Here we make use of the most recent and comprehensive lensed analysis in the HFF from B22, and the field based measurements in B21. Both sets of measurements largely agree with the UVLFs found by other groups across these redshift and magnitude ranges, and we refer the reader to B21, B22, and \cite{Bouwens_2022_I} for detailed comparisons with other related works. 
Regarding the systematic concerns mentioned above, a few key points
are as follows. First, recent studies have found that faint-end source sizes are smaller than previously expected \citep{Bouwens17a} and so minor completeness corrections are adequate in estimating the faint-end UVLF. Second, B22 use median magnification maps compiled from a suite of lensing models for each HFF cluster (e.g. \citealt{Livermore17,Bouwens17b}). Furthermore,
B22 adopt a forward modeling approach which can account for at least much of the uncertainty in the lensing models \citep{Bouwens17b,Atek18}. Specifically, the authors construct mock UVLF data with one lensing model as input and analyze the maps using alternative magnification models. This procedure allows them to assess the UVLF error budget owing to uncertainties in the lensing models.\footnote{Note, however, that we use the binned UVLF estimates from that study, which do not account directly for the uncertainties in the magnification maps. The magnification uncertainties are only included in the error budget for the alternate parametric fit in that work. However, we discuss the impact of uncertainties in the lensed UVLF results further in \S \ref{sec:additional}.} Finally, an encouraging
result is that B22 demonstrates consistency between the faint-end slopes of the UVLF estimates behind lensing clusters and in the field.

Our following analyses make use only of measurements in redshift bins centered at or above $z=5$, i.e. we consider galaxy populations during or slightly after reionization. 
The sample of B21 UVLF measurements used originate from 5,405 galaxies in redshift bins of width $\Delta z=1$ with bin centers spanning $z=5-10$, where the data centered on $z=10$ comes from \cite{Oesch_2018}. These measurements typically cover absolute magnitudes of $M_{\rm UV} = -24$ to $-17$. The B22 lensed UVLF sample used has bin centers at $z=5-9$ of widths $\Delta z=1$ and is comprised of 525 galaxies with absolute magnitudes of about $M_{\rm UV} = -19$ to $-13$.

In order to account for both cosmic variance and systematic uncertainties in the lensed UVLF results, B22 recommend incorporating an additional $\sim20\%$ normalization uncertainty on their lensed UVLF estimates.\footnote{Their recommended uncertainty increases to $\sim22\%$ in the $z=5$ bin where fewer HFF clusters are included in the analysis.} In most of what follows, we neglect this uncertainty, but return to address its possible impact in \S \ref{sec:additional}. 

\section{Methodology}
\label{sec:Method}

Our methodology follows a range of previous modeling efforts which connect UV luminous galaxies to their host dark matter halos (e.g. \citealt{Vale:2004yt,Behroozi:2014tna,Mason:2015cna,Schive:2015kza,Sun2016,Furlanetto2017,Mirocha2023}). Of these works, our modeling is closest to that of \cite{Sun2016} and \cite{Schive:2015kza}.
As mentioned in the Introduction, these approaches are motivated by the notion that galaxy formation is driven by the gravitational collapse of dark matter halos, with gas subsequently falling into dark matter potential wells, cooling, condensing into the halo center, and fragmenting to form stars \citep{White78}. This process is, in turn, regulated by poorly understood feedback effects involving supernova explosions, active galactic nuclei (AGN), and photoionization, among other processes (see e.g. \citealt{Silk2011} for a brief review). Given this context, our semi-empirical modeling involves several key inputs and assumptions. Specifically, we: i) adopt a fitting formula for the halo mass function \citep{Sheth:2001dp}, as derived from N-body simulations (and inspired by analytic calculations), ii) assume that the star formation rate (SFR) in a halo of mass $M$ is proportional to the average baryonic accretion rate onto such halos, with the accretion rate determined from numerical simulations of cosmological structure formation \citep{McBride2009,Springel2005}, and iii) assume a conversion factor relating SFR and UV luminosity.  
The baryonic accretion rate and SFR are related by the SFE as a function of host halo mass $M$ and redshift $z$, $f_\star(M,z)$.\footnote{Sometimes the star formation efficiency is defined instead by the ratio of the stellar mass in a halo to the total baryonic mass of the halo. We denote this alternative quantity by $\tilde{f}_\star(M) = M_\star/(M \Omega_b/\Omega_m)$ as in e.g. \cite{Furlanetto2017}. See also \S \ref{sec:sfe_results}.} Given our assumptions, knowing this relation allows one to model how UV luminous galaxies populate dark matter halos.

In practice, our semi-empirical approach assumes a parametric description for the average $f_\star(M,z)$ motivated by the nature of feedback effects on star formation. While allowing for some scatter in the relation our model calibrates the SFE from UVLF measurements. This model then allows predictions beyond the range in redshift, luminosity, and halo mass in which it is calibrated. This is valuable for understanding the effects of feedback on galaxy formation as well as for determining the properties of the sources of reionization and related topics. This section lays out the main ingredients of this methodology in more detail, including the halo mass function and accretion rate models (\S \ref{sec:hmf}), 
the connection between SFR, SFE, and UV luminosity (\S \ref{sec:sf_efficiency}), 
the conditional luminosity function approach to map between UV luminosity and halo mass (\S \ref{sec:clf}), and alternative models (\S \ref{sec:flat}). In addition, it is important to account for the attenuation of each galaxy's intrinsic UV luminosity by dust grains (\S \ref{sec:dust}). The final step in our methodology is to compare models with the UVLF data: here we use the Monte-Carlo Markov Chain (MCMC) technique to efficiently sample the posterior distribution across the multi-dimensional parameter space of interest (\S \ref{sec:mcmc}). 
\subsection{Halo Mass Function and Accretion Rate Models}
\label{sec:hmf}

The first key ingredient in our methodology is a model for the halo mass function. The halo mass function, denoted here by $n(M)$,
describes the abundance of dark matter halos per unit mass interval such that $n(M) {\rm d}M$ is the number of halos per co-moving volume with mass between $M$ and $M + {\rm d}M$. We adopt the
Sheth-Tormen halo mass function model \citep{Sheth:2001dp}, in which:
\begin{equation}
    n(M) = \frac{\bar\rho_m}{M^2}f_{\rm ST}(\sigma)\left|\frac{{\rm d}\ln\sigma^{-1}}{{\rm d}\ln M}\right|. \label{eq:ST}
\end{equation}
Here $\bar \rho_m$ is the mean matter density per co-moving volume and $\sigma=\sigma(M,z)$ is the root-mean-square (rms) density fluctuation, according
to linear theory, within a top-hat sphere enclosing a mass $M$ at the cosmic mean co-moving matter density. That is, the radius of the top-hat is $R = [3 M/(4 \pi \bar \rho_m)]^{1/3}$. In the Sheth-Tormen model, the function $f_{\rm ST}(\sigma)$ is determined uniquely by $\sigma(M,z)$ relative to the threshold for spherical collapse in linear theory,
$\delta_c$:
\begin{equation}
    \begin{split}
    f_{\rm ST}(\sigma) =& \sqrt{\frac{2a_{\rm ST}}{\pi}}A_{\rm ST}\left[1 + \left(\frac{\sigma^2}{a_{\rm ST}\delta_c^2}\right)^{p_{\rm ST}}\right] \\
    & \times \frac{\delta_c}{\sigma} {\rm exp} \left[-\frac{a_{\rm ST}\delta_c^2}{2\sigma^2}\right].
    \label{eq:f_st}
    \end{split}
\end{equation}
Here we consider the rms density fluctuations at redshift $z$, 
$\sigma(M,z)$, and $\delta_c=1.686$ is the spherical collapse threshold (at the redshift of interest).\footnote{Alternatively, and equivalently, one can extrapolate both the rms density fluctuation and the collapse threshold to $z=0$ using linear theory.} We drop the redshift dependence in our notation for brevity.
The Sheth-Tormen mass function model is motivated by the excursion set formalism \citep{bond_etal1991} and the dynamics of ellipsoidal collapse. The parameter values here, $a_{\rm ST}=0.707$, and $p_{\rm ST}=0.3$, provide good fits to the halo mass function measured from N-body simulations, and $A_{\rm ST}=0.322$ is
a normalization constant \citep{Sheth:2001dp}. Although further simulation tests show that Sheth-Tomen is accurate at high redshifts and low masses, it may over-predict the abundance of rare massive halos at high redshift (see \citealt{Lukic:2007fc} for details). In any case, the differences with more accurate models are likely small compared to the uncertainties in the UV luminosity-halo mass relation and the dust corrections, and so we use the Sheth-Tormen model throughout.   

The next input to our model is the baryonic accretion rate onto dark matter halos. Following \citet{Sun2016}, we adopt the fitting formula for the matter accretion rate from \cite{McBride2009}. Those authors use the Millenium simulation \citep{Springel2005} dark matter halo catalogs to characterize the average mass accretion rate onto halos as a function of their mass and redshift. The average accretion rate was found to follow a simple scaling relation with  
$\dot{M} \propto M^\delta (1+z)^\eta$ at high redshift ($z \geq 1$) with $\delta = 1.127$ and $\eta=2.5$ \citep{McBride2009,Trac:2015rva}. Although this relation was only calibrated on simulated halos with $M \gtrsim 10^{12} {\rm M}_\odot$ and $z \leq 6$, we will assume that it also applies at higher redshift and in smaller mass halos. Further supposing that the infalling matter has the universal baryon fraction of $\Omega_b/\Omega_m$, the {\em baryonic accretion} rate may be written as \citep{Sun2016}:
\begin{equation}
    \dot M_{\rm acc}(M,z) \approx 3 {\rm M}_\odot{\rm /yr} \left(\frac{M}{10^{10}{\rm M}_\odot}\right)^{\delta}\left(\frac{1+z}{7}\right)^\eta \label{eq:M_acc}
\end{equation}
with $\delta=1.127$, $\eta=2.5$, and $M$ is the total (dark matter plus baryons) halo mass.

\subsection{Connection to the Star Formation Efficiency}
\label{sec:sf_efficiency}
We can then connect the baryonic accretion rate with the SFR, provided that some fraction of the accreted baryons are rapidly 
converted into stars. Specifically, denoting this fraction, the average star formation efficiency,
in a halo of mass $M$ at redshift $z$, by $f_\star(M,z)$, we can write:
\begin{equation}
{\rm SFR}(M,z) = f_\star(M,z) \dot M_{\rm acc}(M,z),
\label{eq:sfr_efficiency}
\end{equation}

In practice, we aim to calibrate $f_\star(M,z)$ using the UVLF measurements and so need to consider also the relationship between UV luminosity and SFR. Since most
UV photons are produced by young massive stars, the UV luminosity of a galaxy is generally a good tracer of star formation.
In detail, however, this connection depends on the stellar initial mass function (IMF), the stellar metallicity, stellar binarity, rotation, and the age of the stellar populations \citep{Madau:2014bja}. Here, as in \cite{Sun2016,Madau:2014bja}, we assume a constant conversion factor of:
\begin{equation}
    \text{SFR}(M,z) = \kappa_{\rm UV} L (M,z)\label{eq:SFR_LUV},
\end{equation}
with $\kappa_{\rm UV} = 1.15 \times 10^{-28}$ $\text{M}_\odot$ ${\rm yr}^{-1}$$/$${\rm erg}$ ${\rm s}^{-1}$ ${\rm Hz}^{-1}$. Here $L(M,z)$ is the specific luminosity at a rest-frame wavelength of $1500 \, \angstrom$ (before attenuation by dust) in units of ${\rm erg}$ ${\rm s}^{-1}$ ${\rm Hz}^{-1}$.\footnote{Note that the B21 and B22 results are given at $1600 \, \angstrom$ in the rest-frame as opposed to $1500 \, \angstrom$ but the conversion factor is insensitive to this difference \citep{Madau:2014bja}.}
As discussed in \cite{Furlanetto2017}, this conversion factor is
likely uncertain at the factor of a few level owing to the
dependencies mentioned above. As long as $\kappa_{\rm UV}$ 
does not evolve strongly across the redshift range considered here, $z \sim 5-10$, and/or depend significantly on halo mass, we expect the uncertainties in $\kappa_{\rm UV}$ to mainly impact the overall normalization of $f_\star(M,z)$ rather than the inferred trends with redshift and halo mass. 

Although for brevity of notation we have not made it explicit here, the key quantities discussed in this section: $\dot M_{\rm acc}$, SFR, $f_\star$, and $L$, should be thought of, respectively, as the the {\em average} accretion rate, SFR, SFE, and UV luminosity in a halo of mass $M$ at redshift $z$. We will account for {\em scatter} in the UV luminosity-halo mass relation when we connect the observable UVLF and the model halo mass function, as described below. 

\subsection{The Conditional Luminosity Function}
\label{sec:clf}

In practice, we work backwards to infer $f_\star(M,z)$ from the data: given the UVLF measurements (\S \ref{sec:uvlf_data}) and the halo mass function model (Equations \ref{eq:ST}-\ref{eq:f_st}), we
determine the mapping between UV luminosity and halo mass, while the average SFE then follows from Equations \ref{eq:M_acc} - \ref{eq:SFR_LUV}. In order to relate UV luminosity and halo mass we use
a conditional luminosity function (CLF) model \citep{Yang2003,Cooray05,Bouwens2015,Schive:2015kza}.

The CLF quantifies the link between the halo mass function, $n(M,z)$, and the UV luminosity function, $\phi(L,z)$. Here
$\phi(L,z) {\rm d} L$ is the number of galaxies per co-moving volume with specific UV luminosity between $L$ and $L + {\rm d}L$ at redshift $z$.
In the fiducial CLF model adopted here, we suppose
that each host dark matter halo (above some limiting mass $M_{\rm min}$) houses
a single UV luminous galaxy at the center of the halo. That is, we neglect any contribution from satellite galaxies as current $z \gtrsim 5$ clustering measurements suggest satellite fractions smaller than a few percent \citep{Harikane_2022}. We further assume a ``duty cycle'' of unity so that all halos above $M_{\rm min}$ host
UV luminous galaxies at any given time. We explore alternative scenarios in \S \ref{sec:alt_models}. 

Explicitly, the CLF $\phi_c(L|M,z)$ links the halo mass function and the luminosity function via:
\begin{equation}
    \phi(L,z) = \int_{M_{\rm min}}^\infty  \phi_c(L|M,z) n(M, z) \dif M. \label{eq:phi}
\end{equation} 
Throughout most of this work we consider $M_{\rm min} = 10^8 {\rm M}_\odot$, comparable to the atomic cooling mass \citep{barkana_loeb2001}. We comment on the impact of this assumption in what follows when relevant. 
Put differently, the quantity $\phi_c(L|M,z)$ specifies the conditional probability that a halo of mass $M$ at redshift $z$ hosts a galaxy with specific UV luminosity $L$. 

Following previous work, we suppose that the CLF follows a lognormal distribution, parameterized by the median galaxy luminosity in a halo of mass $M$ (and redshift $z$), $L_c(M,z)$, and the scatter around this relation, $\sigma$ \citep{Cooray05,Bouwens2015,Schive:2015kza}.
Hence,

\begin{equation}
    \phi_c(L|M,z) = \frac{1}{\sqrt{2\pi}\sigma L} \exp{-\frac{\ln\left[L/ L_c(M,z)\right]^2}{2\sigma^2}}. \label{eq:CLF}
\end{equation} 

In our fiducial model, we further follow previous work and adopt a five parameter description of the relationship between the median UV luminosity and halo mass. This description is intended to capture a range of possibilities for the feedback-regulated dependence of the SFE, and hence UV luminosity, on halo mass, and the redshift evolution in this relation. Specifically, as in \cite{Schive:2015kza}, we take:
\begin{equation}
    L_c(M,z) = L_0\left[\frac{(M/M_1)^p}{1+(M/M_1)^q}\right]\left(\frac{1+z}{7}\right)^r, \label{eq:L_c}
\end{equation}
with $L_0$ and $M_1$ controlling the normalization of the relationship, $p$ and $p-q$ the power-law scalings with mass for $M<<M_1$ and $M>>M_1$ respectively, and $r$ the redshift dependence. 

Our fiducial model adopts a mass independent value of the scatter (in the natural logarithm of the UV luminosity), $\sigma = 0.37$, equivalent to $0.16$ dex of scatter \citep{Schive:2015kza}. Note that the {\em shape} of the UV luminosity-halo mass relationship is assumed to be redshift independent in Equation \ref{eq:L_c}. This is a simplifying assumption
motivated in part by the limited redshift range probed in our analysis.

This functional form has been found to provide a good fit to previous UVLF measurements \citep{Bouwens2015,Schive:2015kza}. As shown in more detail in what follows, this parameterization describes cases where the SFE peaks in the general vicinity of $M \sim M_1$ and declines towards both smaller and larger masses. Physically, this 
behavior may reflect the combined influence of supernova/photoionization feedback at low masses and AGN feedback at large mass scales. A key motivation of the current work is to test whether this description still applies at the small luminosities probed by the B22 HFF results, where feedback effects may be especially strong. We will also explore alternative contrasting models in the low luminosity regime 
(\S \ref{sec:flat}, \S \ref{sec:alt_models}).

Finally, for completeness note that we often work in units of absolute magnitude, $M_{\rm UV}$ \citep{oke74}, such that \begin{equation}
    M_{\rm UV} = 51.6 - 2.5\log_{10}\left(L/\left[{\rm erg}\ {\rm s}^{-1}\ {\rm Hz}^{-1}\right]\right). \label{eq:M_to_L}
\end{equation}
Now $\phi(M_{\rm UV},z)$ denotes the abundance of galaxies per unit UV magnitude, as a function of UV magnitude, $M_{\rm UV}$, whereas $\phi(L,z)$ gives the abundance per unit luminosity. The mapping between these luminosity functions is:
\begin{equation}
    \phi(M_{\rm UV},z) = \phi\left(L(M_{\rm UV}),z\right) \left|\frac{\dif L}{\dif M_{\rm UV}}\right|.  \label{eq:phi_m2}
\end{equation}

\subsection{Alternative Models Considered}
\label{sec:flat}

In addition to our fiducial CLF model of Equations \ref{eq:phi}-\ref{eq:L_c}, we consider several potential model variants, many of which imply enhanced SFE relative to our fiducial model at low mass scales.
This is motivated partly by the slight, yet systematic, underprediction of the lensed UVLF measurements in our fiducial model. 
These cases will be discussed further in \S \ref{sec:alt_models}. However in the proceeding sections we will often consider a ``flattening model'' as a simple illustrative alternative to compare with our fiducial results. In the flattening model, the SFE is assumed to be a constant function of halo mass between $M_{\rm min}$ and a mass scale set by an additional free parameter $M_2$ (see also \citealt{Sun2016}, their ``Model I").
In \S \ref{sec:alt_models}, we consider further possibilities including models where the SFE is strongly truncated below some mass scale, the scatter in UV luminosity is enhanced at small masses, or where the duty cycle of star formation activity departs from unity.

\subsection{Dust Correction}
\label{sec:dust}

The UV luminous galaxies considered here may harbor substantial amounts of dust, especially at the bright end of the luminosity function, and this will attenuate the galaxies' UV luminosities. We will follow previous work (e.g. \citealt{Smit2012,Sun2016,Sabti2022}) in applying an average ``correction'' to account for dust extinction: that is, we will estimate the average dust extinction in each UV magnitude bin of B21 and B22 and use this to determine the ``intrinsic'' UVLF (and revised error bars) with the impact of dust attenuation removed. We then compare models with the intrinsic UVLF throughout. 

We briefly review the dust attenuation correction here, and refer the reader to the literature (e.g. \citealt{Smit2012,Vogelsberger2020,Sabti2022}) for further details and discussion. The dust attenuation estimate makes use of the fact that the portion of the UV emission from star-forming galaxies which is absorbed by dust is then re-radiated in the infrared. The ratio of the far-infrared to observed UV flux is hence an indicator of the amount of UV flux absorbed by dust. At least in the more local universe, this flux ratio -- termed IRX for infrared excess -- (IRX $\equiv L_{\rm IR}/L$) correlates with the observed UV continuum spectrum of the star-forming galaxies \citep{Meurer99}. In other words, the UV spectral slope $\beta$ (with specific flux $f_\lambda \propto \lambda^\beta$) correlates with the dust attenuation, $A_{\rm UV}$, where this
latter quantity describes the number of magnitudes of dust attenuation. Hence measurements of the spectral slope $\beta$ may be used to estimate the intrinsic UV emission from the observed UV luminosity \citep{Meurer99,Smit2012}.

In order to determine the average attenuation in a given UV luminosity (or $M_{\rm UV}$) bin, we require the mean spectral slope, $\langle \beta \rangle$, at the luminosity in question and the scatter, $\sigma_\beta$. Suppose that the observed dust attenuated luminosity is $L_{\rm obs}$, while the average intrinsic UV luminosity is simply $\langle L \rangle$ (note the brevity of notation, both are specific luminosities), then the average attenuation is:
\begin{equation}
\langle A_{\rm UV} \rangle \equiv 2.5 \, {\rm log}_{10}\left( \frac{\langle L \rangle}{L_{\rm obs}}\right).
\label{eq:dust_avg_def}
\end{equation}
If we then assume a linear correlation with $A_{\rm UV} = C_0 + C_1 \beta$, the average of Equation \ref{eq:dust_avg_def} can be calculated assuming a Gaussian distribution of spectral slopes $\beta$. Specifically, the average attenuation is \citep{Smit2012,Vogelsberger2020,Sabti2022}:
\begin{equation}
    \langle A_{\rm UV}\rangle\left(z, M_{\rm UV}\right)= C_0 + 0.2\ln(10)\sigma_\beta^2 C_1^2 + C_1\langle \beta \rangle(z, M_{\rm UV}), \label{eq:dust_A}
\end{equation}
where $\langle \beta \rangle$ varies with UV magnitude and redshift. Equation \ref{eq:dust_A} applies as long as $\langle A_{\rm UV} \rangle \geq 0$, otherwise we set it to zero. 
The dust attenuation thus depends on the parameters $C_0$ and $C_1$ which describe the correlation between $A_{\rm UV}$ and $\beta$ \citep{Meurer99}, while $\langle \beta \rangle$ and $\sigma_\beta$ characterize the distribution of spectral slopes.
We adopt the values $C_0 = 4.54$, $C_1 = 2.07$, $\sigma_\beta=0.34$ \citep{Sabti2022,Overzier11}, and the redshift dependent form of $\langle \beta \rangle$ for $z=5-8$ from \cite{Bouwens:2013hxa}.  
We neglect any dust correction in the redshift bins with $z \geq 9$, where there is too little data to make a robust estimate of dust attenuation. This could lead to a slight underprediction of the SFE at higher redshifts.

However, for $z=5-8$, the estimate of Equation \ref{eq:dust_A} already predicts no dust correction is needed for UV magnitudes fainter than $M_{\rm UV} \sim -17$. As dust attenuation is less important at higher redshifts, neglecting its effects at $z=9-10$ should not influence our conclusions regarding the faint end of the UVLF. At the bright end the dust correction can have a relatively strong effect: in the brightest observed bins, $\langle A_{\rm UV} \rangle\sim1.5-2$ for $z=5-7$, but the correction is smaller with $\langle A_{\rm UV} \rangle\sim0.5-1$ by $z=8$. 

In order to determine the intrinsic UVLF, we need to apply
a correction based on $\langle A_{\rm UV} \rangle$ to each of: the observed UV magnitudes at the center of the bins used in the UVLF estimates, to the UV magnitude bin-widths, to the UVLF estimates $\phi(M_{\rm UV},z)$, and to the measurement errors, as in Equations 3.4-3.7 of \cite{Sabti2022}. 

\subsection{Model Fitting and Parameter Estimation}
\label{sec:mcmc}
One of our main goals is to obtain confidence intervals in the multi-dimensional parameter space describing the UV luminosity-halo mass relation (Equation \ref{eq:L_c}). This may be characterized by a parameter vector ${\bm \theta} = \left(p,q,r,M_{\rm UV,0},{\rm log_{10}}(M_1/{\rm M}_\odot)\right)$. The data vector ${\bf d}$ in this case describes the dust-corrected UVLF measurements
in different magnitude and redshift bins. Typically, we consider the joint fit to the combination of the B21 and B22 measurements across six redshift bins from $z=5-10$. We assume that the B21 and B22 measurements are entirely independent of each other and simply combine them.   
We will also consider the fit to B21 alone (i.e. including only data in the field and ignoring the lensed UVLF measurements from the HFFs). In this case, the data vector is modified accordingly. 
The posterior probability $p({\bm \theta}|{\bf d})$ follows from Bayes Theorem as 
$p({\bm \theta}|{\bf d}) \propto p({\bf d}|{\bm \theta}) p({\bm \theta})$, where $p({\bf d}|{\bm \theta})$ describes the likelihood
of the data given the model parameters, and $p({\bm \theta})$ characterizes the prior probabilities on the parameters. 

We adopt uniform priors on each parameter over the following range $\bm \theta=\left(p,q,r,M_{\rm UV,0}, {\rm log_{10}}(M_1/{\rm M}_\odot)\right)$: $([0.8, 2.5], [0.8, 2.5], [0, 2.5], [-25, -20], [10, 14])$. This prior range was adopted based on previous CLF fits to earlier UVLF data from \cite{Bouwens2015,Schive:2015kza}. These span a conservative range in parameter values and ultimately the likelihood functions are well-peaked within the range spanned by the priors. We hence expect the results to be insensitive to the precise choice of priors here. 

For the most part, we assume a Gaussian likelihood function such that (minus twice) the natural logarithm of the likelihood function (or equivalently the $\chi^2$ value) may be written as:
\begin{equation}
    \chi^2 = - 2 {\rm ln} \left[p({\bf d}|{\bm \theta})\right] = 
    \sum_i \frac{\left[\phi_{\rm model}(i) - \phi_{\rm data}(i)\right]^2}{\sigma^2(i)},
    \label{eq:chisq}
\end{equation}
where the sum over $i$ runs over both different magnitude and redshift bins. The model UVLF, $\phi_{\rm model}(i)$, is determined at the center of the corresponding absolute magnitude and redshift bin using Equations \ref{eq:ST}, \ref{eq:phi}-\ref{eq:phi_m2}. We assume that the measurement errors, $\sigma(i)$, in different magnitude and redshift bins are uncorrelated. 

We incorporate two small refinements to the treatment of Equation \ref{eq:chisq} in order to account for the sometimes asymmetric UVLF measurement errors. Specifically, some UVLF points include only upper limits on the UVLF in a given magnitude bin. In this case, we assume a half-Gaussian likelihood function.
In other bins, the error bars are asymmetric with an upper error of $\sigma_+$ and lower error of $\sigma_-$. In this case, we adopt the treatment of \cite{barlow2004asymmetric} following their Equations 13-16. 

In order to sample the posterior probability distribution and obtain confidence intervals (given the priors, $p({\bm \theta})$, and the likelihood function, $p({\bf d}|{\bm \theta})$, Equation \ref{eq:chisq}), we use the MCMC sampler from the \texttt{emcee} package \citep{Foreman-Mackey12}.
In our \texttt{emcee} runs, we start by initializing 96 walkers in a Gaussian ball with a dispersion of $\sigma_{\rm init}=0.1$ around the best-fit parameters found in Appendix A of \cite{Bouwens2015}. We run for over 10,000 iterations in our fiducial model and have cross-checked that the results are stable after doubling the number of iterations. Additionally, 
given that the longest auto-correlation time is $\tau \sim 70$, we remove the first $\sim 1,000$
samples. This removes the influence of the initialization steps, which fades after about $\sim 10 \tau$ \citep{Foreman-Mackey12}. Some of the alternative scenarios in \S \ref{sec:alt_models} require further iterations, and we have verified convergence in those cases as well. We use the \texttt{scipy.optimize} \citep{2020SciPy-NMeth} module to find the global best fit  parameters (i.e., those which minimize $\chi^2$.)

\section{Results \& Discussion}
\label{sec:results}

\subsection{Conditional Luminosity Fits}
\label{sec:clf_fits}

\begin{figure*}[t]
  \centering
  \includegraphics[width=\textwidth]{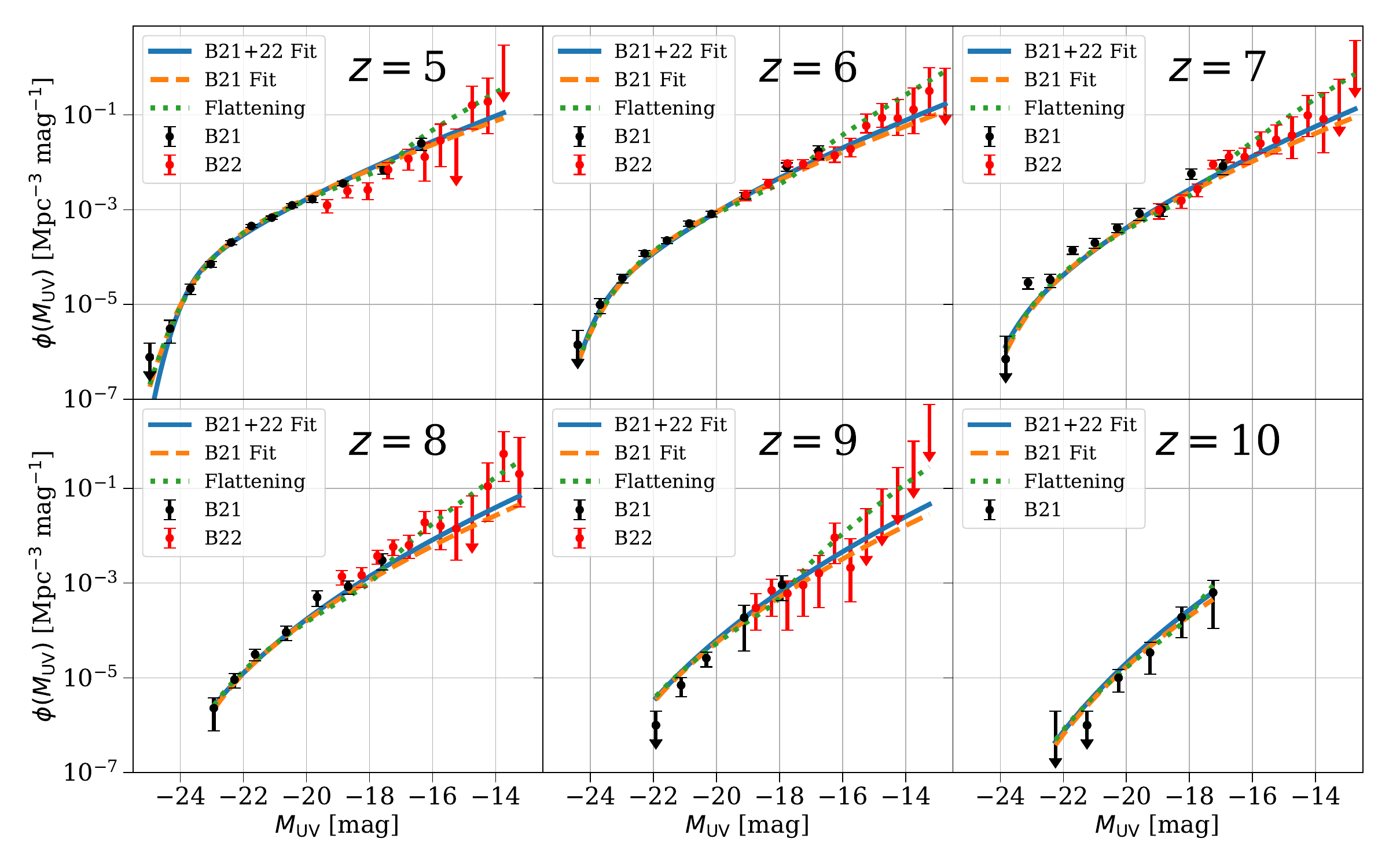}
  \caption{
  Dust-corrected UVLF measurements and $1-\sigma$ error bars (or upper limits) from $z=5-10$ compared with best-fits from three of our models. Measurements from B21 are in black and B22 in red. The solid blue curve is our CLF fit (Equations \ref{eq:phi}-\ref{eq:phi_m2})  to the joint B21+22 data set, the dashed orange curve is a CLF fit to only the B21 measurements, and the green dotted curve is a modified CLF fit to B21+22 with a flat SFE at low mass (see \S\ref{sec:flat}, \S \ref{sec:alt_models}). Note that although the lensed UVLF measurements probe up to five magnitudes fainter than in the field, the best fit to the combined B21+B22 data is similar to that from B21 alone. However, the preferred flattening model is steeper at the faint-end. 
  }
\label{fig:LF}
\end{figure*}

Figure \ref{fig:LF} compares the (dust-corrected) measurements from B21 and B22 with our models best fit to either the joint B21+B22 data set or to the B21 measurements alone. This comparison spans a redshift range of $z \sim 5-10$ and covers around 12 magnitudes in UV luminosity in many of the redshift bins.

Across the range of redshifts and luminosities probed by the union of the B21+B22 points, the baseline CLF model described by Equations \ref{eq:phi}-\ref{eq:L_c} provides a fairly good description. Quantitatively, the best fit model to the joint data set has a $\chi^2=177$. With five model parameters and 111 data points, the reduced chi-squared is $\chi^2_\nu=1.67$. The best fit model parameters are ($p=1.69$, $q=1.58$, $r=1.12$, $M_{\rm UV,0}=-23.79$, $M_1=10^{11.93}$ M$_\odot$). Table \ref{table:params} provides best fit parameter values and their corresponding marginalized 68\% confidence intervals along with metrics describing goodness-of-fit for this and our other model fits. Note that our assumption of a redshift invariant shape for the median UV luminosity-halo mass relationship appears to provide a successful description of the current data across $z \sim 5-10$. Interpretations of these preferred parameter values are discussed in \S \ref{sec:params}. 

Interestingly, the combined best fit to B21+B22 is fairly similar to a simple extrapolation from the B21 measurements alone. 
 This is far from a foregone conclusion since the measurements in the HFFs extend down to much smaller luminosities. These low luminosity galaxies likely reside in small mass dark matter halos where the potential wells are shallow and susceptible to feedback: apparently, these effects do not significantly suppress the faint end of the UVLF over the range of luminosities probed (see also \S \ref{sec:alt_models}). In fact, as can be inferred from comparing the blue solid (joint fit) and orange dashed curves (B21 alone) in Figure \ref{fig:LF},
 the combined fit has a slightly steeper faint end slope than the match to B21 alone (i.e. the UVLF appears enhanced rather than suppressed). Note that in the alternative flattening case (green dotted curve), the UVLF is even steeper at the faint end. 
 
For quantitative comparison, the best fit to the B21 points alone yields the parameters ($p=1.84$, $q=1.34$, $r=0.87$, $M_{\rm UV,0}=-23.01$, $M_1=10^{11.66}$ M$_\odot$), with $\chi^2_{\rm B21} =75.2$ and, with five parameters and 50 data points, $\chi^2_{\nu,{\rm B21}}=1.67$. If these same parameters are compared with the joint B21+B22 data set, then the goodness-of-fit values become $\chi^2=198$ and $\chi^2_\nu=1.86$. 
That is, the best fit parameters do shift and
the old values are comparatively disfavored but not hugely so. 
Specifically, approximating the parameter errors from both cases as independent and adding them in quadrature, the shift in the best fit value of $p$ is just under $2-\sigma$ and is less in the other parameters.
In particular, including the B22 data leads to a smaller $p$, which corresponds to a slightly steeper faint-end UVLF slope. The parameter constraints do, however, tighten significantly after including the B22 lensed data with the error bars on $p$,
$M_{\rm UV,0}$, and ${\rm log_{10}}(M_1/{\rm M}_\odot)$ all shrinking by about a factor of $2-3$. The increased lever arm in luminosity provided by the lensed UVLF data strongly sharpens the constraints on the faint end parameter $p$, which also helps tighten the confidence intervals on other model parameters by breaking degeneracies. 

\subsection{Posterior Distributions and Parameter Degeneracies}

\begin{figure*}[t]
  \centering
    \begin{overpic}[width=\textwidth]{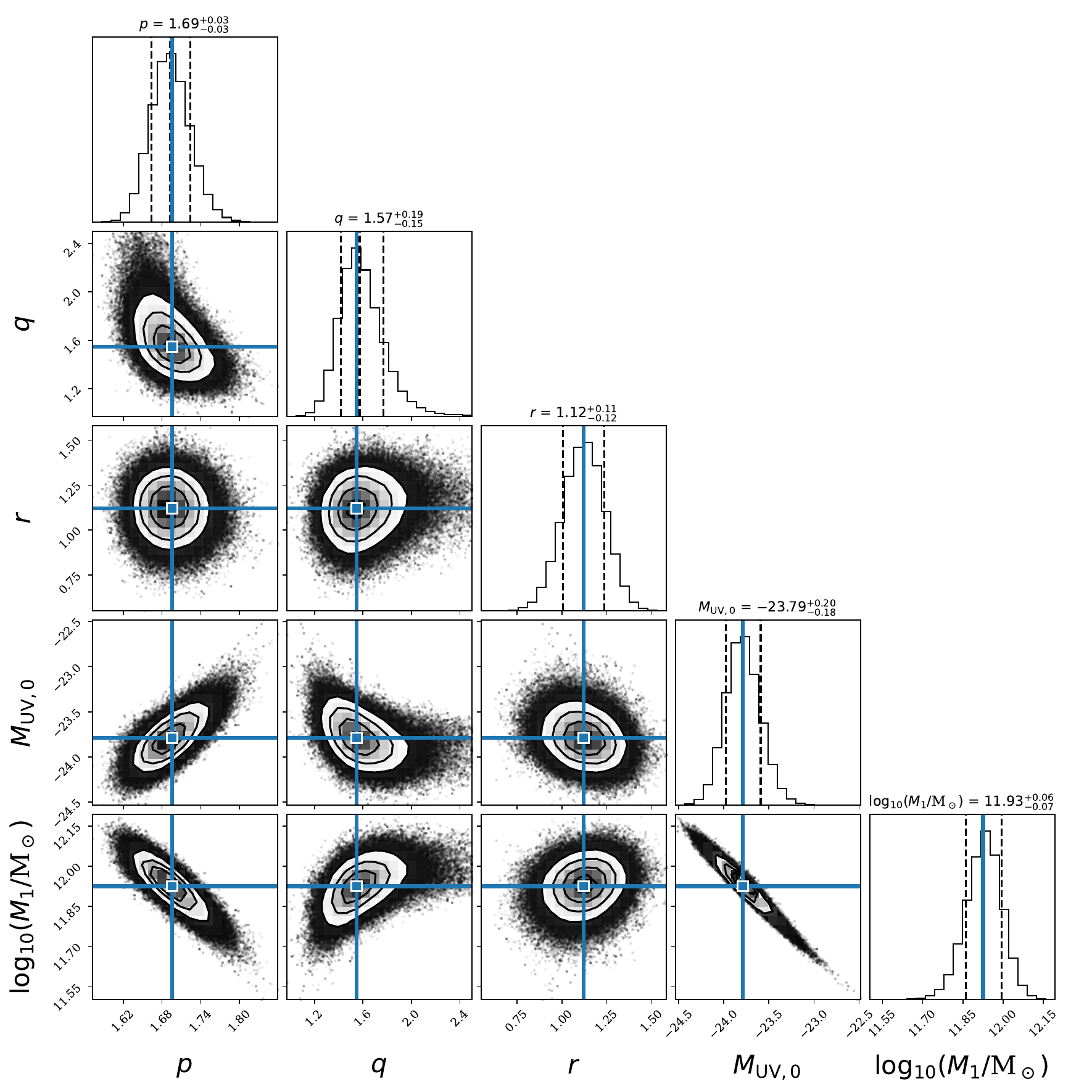}
    \put(45,65){\includegraphics[width=0.65\textwidth]{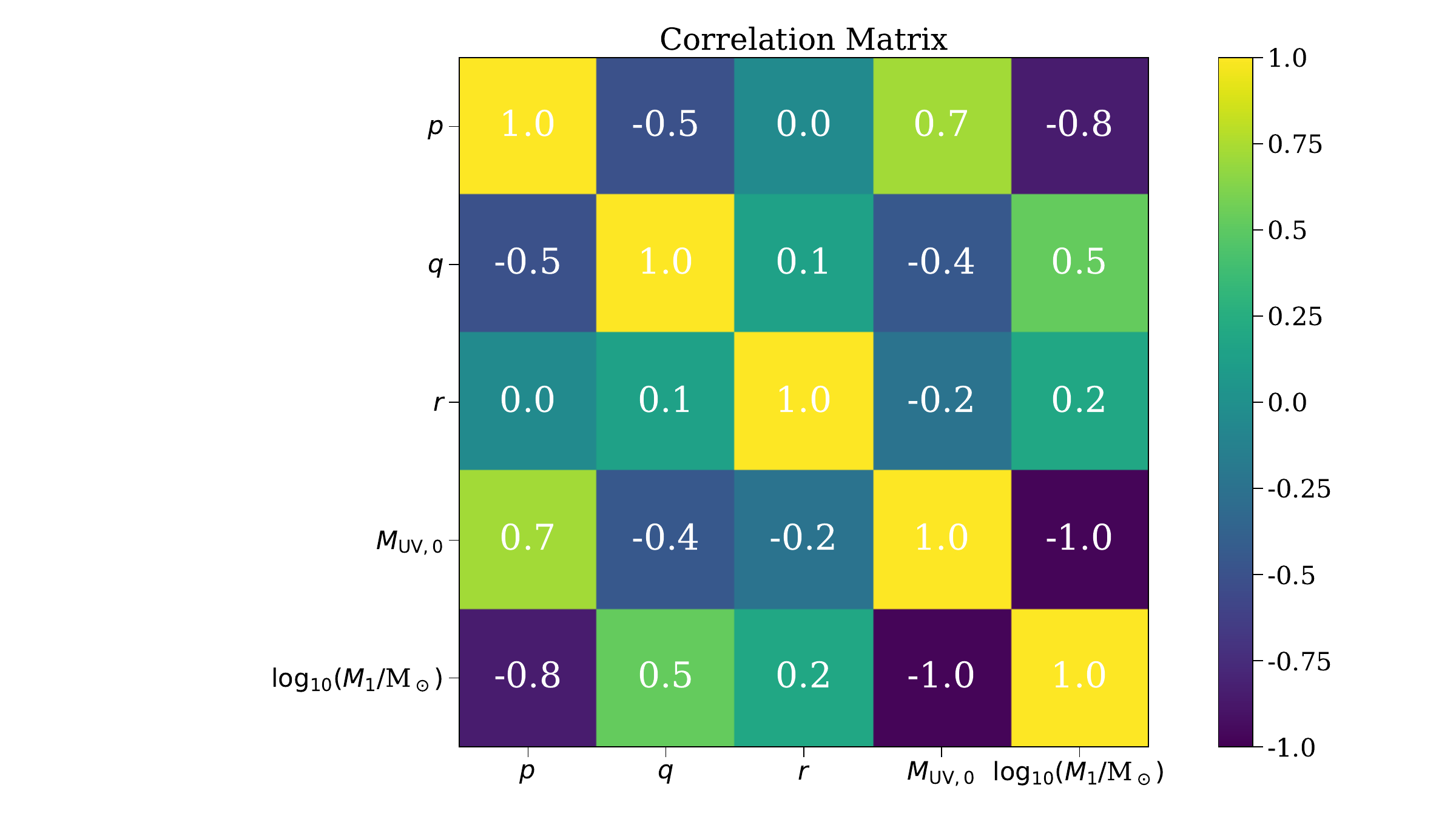}}
    \end{overpic}
  \caption{Posterior distributions in the five-dimensional parameter space of our CLF model fits from \texttt{emcee}. See Equation \ref{eq:L_c} and the accompanying text for a description of the model parameters, which describe the trend of the median UV luminosity with halo mass. 
  The dashed black lines in the histogram panels indicate the 1-$\sigma$ bounds on each parameter. The contours show
  the $\sim$12, 39, 68, and 86\% confidence intervals in each 2D plane. The blue lines and squares show the best fit parameter values from \texttt{scipy.optimize}.  The table in the upper right hand corner gives the correlation coefficients between the different parameter constraints. See the text for further discussion regarding the best fit parameter values and degeneracy directions.
    }
  \label{fig:mcmc}
\end{figure*}

Further, Figure \ref{fig:mcmc} displays the posterior distributions for our five-parameter fits to the B21+B22 data from \texttt{emcee}, showcasing the marginalized $68\%$ confidence intervals on each parameter and illustrating the covariance between different parameters. 
Each of the model parameters are well constrained by the combined data, with fractional errors of $\sim 10\%$ on
$q$ and $r$, while the other parameters are determined to better than $\sim 5\%$. In all 
cases, the posterior distributions are narrow and lie within the range of the input priors (\S \ref{sec:mcmc}).
The larger relative errors on $q$ and $r$ arise because these parameters control the bright-end
luminosity halo mass relation and the redshift evolution, respectively. The galaxies at the bright end, controlling $q$, are rare objects and so estimates of their abundance are subject to large Poisson uncertainties. Our current comparison with the UVLF data spans a relatively small range of redshifts, $z \sim 5-10$, and so this gives a limited handle on the redshift evolution parameter, which is likely related to the larger errors on $r$. 

Our conclusions regarding the star formation efficiency and reionization history are somewhat dependent on inferences based on each parameter separately. Thus, although the parameters are well-constrained, it is useful to keep in mind any degeneracies between parameters. As shown in Figure \ref{fig:mcmc}, there exists a very strong negative correlation between $M_{UV,0}$ and $M_1$, as well as moderate correlations between each of these parameters and $p$ and $q$. This is mainly a consequence of the measurements probing the low-mass limit of the double power law $L(M)$ relation much more strongly, where one would expect a perfect degeneracy between $M_{UV,0}$, $M_1$ and $p$. Additionally, $q$ only modulates the bright-end slope as part of $p-q$.

\begin{table*}[t]
\centering
\begin{tabular}{@{}lllllllllll@{}}
\toprule
Model       & $p$  & $q$  & $r$  & $M_{\text{UV},0}$  & $\log_{10} \left(\frac{M_1}{{\rm M}_\odot}\right)$ & $\log_{10} \left(\frac{M_2}{{\rm M}_\odot}\right)$ & Added Param & $\chi^2$ & $\chi^2_\nu$ & $\Delta$AIC \\ \midrule
B21+22      & $1.69\pm^{0.03}_{0.03}$ & $1.58\pm^{{0.20}}_{0.16}$ & $1.12\pm^{0.11}_{0.12}$ & ${-23.79}\pm^{{0.20}}_{{0.19}}$ & $11.93\pm^{0.06}_{0.07}$ & $-$ & $-$ & $177$ & $1.67$ & $24.3$ \\

B21         & $1.84\pm^{{0.09}}_{0.07}$ & $1.34\pm^{0.12}_{0.10}$ & $0.87\pm^{0.13}_{0.13}$ & ${-23.01\pm^{0.45}_{0.36}}$ & ${11.66}\pm^{0.12}_{0.14}$ & $-$ & $-$ & $198$ & $1.86$ & $36.3$\\

Flattening  & $1.99\pm_{0.10}^{0.12}$ & ${1.43}\pm_{0.09}^{0.10}$ & $1.08\pm_{0.12}^{0.12}$ & ${-22.71}\pm_{0.40}^{{0.45}}$ & ${11.54}\pm_{0.13}^{0.13}$ & $10.41\pm^{0.06}_{0.07}$ & $-$ & $156$ & $1.49$ & $12.1$ \\

Shallow-slope  & $2.09\pm_{0.12}^{{0.14}}$ & $1.43\pm_{0.09}^{{0.08}}$ & $1.12\pm_{0.12}^{0.12}$ & ${-22.34}\pm_{{0.47}}^{{0.52}}$ & $11.43\pm_{{0.14}}^{0.14}$ & ${10.55}\pm_{{0.05}}^{0.05}$ & $0.27\pm_{0.06}^{{0.06}}$ & $137$ & $1.32$ & $0$ \\

Bursty & ${1.40}\pm_{0.05}^{{0.06}}$ & ${1.78\pm_{0.46}^{1.27}}$ & ${0.96\pm_{0.12}^{0.12}}$ & ${-24.35\pm_{0.22}^{0.32}}$ & ${12.24}\pm_{0.10}^{0.09}$ &  $11.08{\pm_{0.05}^{0.05}}$ & ${0.48\pm_{0.06}^{0.07}}$ & $143$ & $1.38$ & $4.5$ \\

High Scatter  & $1.85\pm_{0.05}^{0.05}$ & $1.46\pm_{0.11}^{0.12}$ & $1.03\pm_{0.13}^{0.12}$ & ${-23.20\pm_{0.26}^{0.28}}$ & $11.71\pm_{0.09}^{0.08}$ & $9.71\pm_{{0.22}}^{0.14}$ & ${1.51}\pm_{0.18}^{0.19}$ & $150$ & $1.44$ & $9.4$ \\

Truncation  & $1.69\pm^{0.03}_{0.03}$ & $1.58\pm^{0.20}_{0.16}$ & $1.12\pm^{0.11}_{0.12}$ & ${-23.97\pm^{0.20}_{0.19}}$ & $11.93\pm^{0.06}_{0.07}$ & $8.68\pm_{0.46}^{0.46}$ & $-$ & $177$ & $1.69$ & $26.3$  \\ \bottomrule
\end{tabular}
\caption{Best-fit model parameters with $68\%$ confidence intervals, $\chi^2$ values, the $\chi^2$ per degree of freedom ($\chi^2_\nu$), and $\Delta$AIC values for our model fits to the B21+22 joint UVLF measurement set (see \S \ref{sec:params}). The first two rows use our fiducial CLF model, but the parameters in the second row are what are preferred by a fit to the B21 measurements alone. The remaining five models are modifications to the CLF described in \S\ref{sec:alt_models} that allow for alternative behavior in halos of mass below $M_2$, a new sixth parameter. In the shallow-slope, bursty, and high scatter models the added seventh parameter is $s$, $f_{\rm duty}$, or $\sigma$, respectively, as described in \S \ref{sec:alt_models} of the text.
}
\label{table:params}
\end{table*}

\subsection{Parameters \& Their Interpretation}
\label{sec:params}

\begin{figure}[ht]
  \centering
  \includegraphics[width=0.45\textwidth]{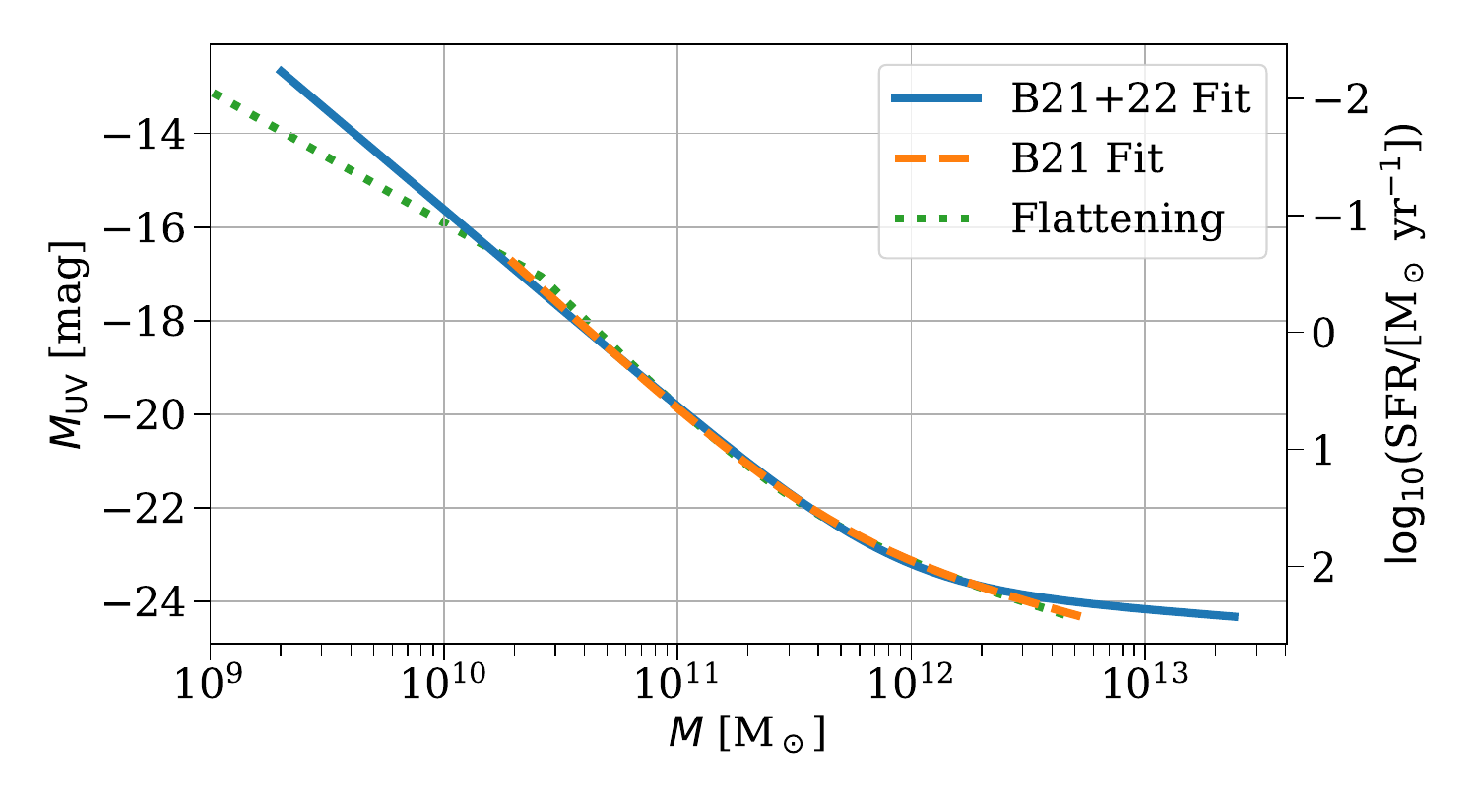}
  \caption{Mean UV luminosity-halo mass relationship implied by the CLF model fits at $z=6$. The line and color styles are similar to those in Figure \ref{fig:LF} and will be used in the figures throughout: the orange dashed curve is our model fit to the B21 measurements alone, while the blue solid curve also includes the B22 lensed UVLF measurements, and the green dotted curve is the alternative flattening scenario for the joint fit to B21+B22. 
  Here we do not extrapolate beyond the brightest and dimmest measurements in the relevant data set. The corresponding star formation rates (from Equation \ref{eq:SFR_LUV}) are shown along the y-axis to the right. The B22 measurements probe down to halo mass scales as small as $M \sim 2 \times 10^9 {\rm M}_\odot$  in our fiducial scenario (with still smaller masses probed in the flattening case),
  about an order of magnitude smaller than possible with the B21 observations alone. 
  }
  \label{fig:lum-mass}
\end{figure}

In order to understand the implications of our best-fit parameters, it is helpful to examine the average UV luminosity-halo mass relationships in our models. This is shown in Figure \ref{fig:lum-mass} at redshift $z=6$ for the joint B21+B22 data fit (solid blue curve), the fit to B21 alone (dashed orange curve), and the flattening fit to B21+22 (dotted green curve). The functional form of $L_c(M,z)$ in our model preserves the shape of this relation across redshift (Equation \ref{eq:L_c}) and the redshift evolution is fairly mild.
For each curve we show only the range of UV luminosity/halo mass scales directly probed by the data at $z=6$ and so the B21 curve spans roughly $M_{\rm UV} \sim -24$ to $-17$, while the fits to B21+B22 extend the reach in UV magnitudes out to $M_{\rm UV} \sim -13$. The corresponding average SFRs, assuming Equation \ref{eq:SFR_LUV}, span
roughly from 
${\rm SFR} \sim 0.01-100 {\rm M}_\odot$ yr$^{-1}$ (as indicated by the y-axis markings on the right-hand side of the figure).  
Notably, the average host halo mass in the best fit model at the faint end of the B22 measurements is around 
$M \sim 2 \times 10^9 {\rm M}_\odot$ (and still smaller in the alternative flattening scenario shown by the green dotted line). This is about an order of magnitude smaller than probed in B21 (according to our best fit models). 
The best fit fiducial parameters to B21+B22 put the brightest galaxies (from B21 measurements, near $M_{\rm UV} \sim -24$) in slightly more massive halos than the fit to B21 alone, and so the blue solid curve extends to slightly larger halo masses than its orange counterpart. As previously mentioned in \S \ref{sec:dust}, we should bear in mind that the bright-end behavior is sensitive to the uncertain dust correction: the intrinsic $z \sim 6$ UV luminosity is brighter by $\sim2-2.5$ magnitudes (i.e. almost an order of magnitude in luminosity) after the dust attenuation adjustment there. 

The curves also reveal the double power-law form of
the mean UV luminosity-halo mass relationship (in the fiducial, non-flattening scenarios),
following $L \propto M^p$ at $M << M_1$ and
$L \propto M^{p-q}$ at $M >> M_1$\footnote{Note that $L\propto M^b \iff M_{\rm UV} \propto -2.5 \, b M$.}. The best fit parameters give a transition mass around $M_1 \sim 4-9 \times 10^{11} {\rm M}_\odot$, with the precise value depending on which data
set we fit to and the model considered (see Table \ref{table:params}). 

The parameter $p$ is also related to the faint-end slope of the UVLF, $\alpha$, (where $\phi(L)\propto L^{-\alpha}$ approximately holds). In the low mass $M\ll M_1$ regime, $L\propto M^p$, and the halo mass function can be approximated as a power law. If $n(M)\propto M^{-(\gamma+1)}$, it then follows from Equation \ref{eq:CLF} that $\phi(L)\propto L^{-(1+\gamma/p)}$. From this relation our fiducial model predicts an approximate faint end slope of $\alpha=1.67$ ($1.75$) at $z=6$ for halos around mass $M=2 \times 10^9 {\rm M}_\odot$ ($M=10^{10} {\rm M}_\odot$). Smaller $p$ corresponds to a steeper UVLF faint-end slope, and for similar reasons a smaller $p-q$ corresponds to a sharper bright-end fall off.

The connection between the other parameters and the UVLF can also be understood simply. Increasing $L_0$ ($M_1$) boosts (lowers) the model UVLF in every magnitude bin. $M_1$ serves as the scale parameter separating the bright- and faint-end behavior of $L(M)$ (which, as we will see, implies that the SFE peaks in halos near $M\sim M_1$), but the location of the exponential fall-off at the bright-end of the UVLF is instead driven almost entirely by the halo mass function. Finally, at a given $L$ increasing $r$ puts higher redshift galaxies in comparatively smaller, yet more numerous, halos. This implies that increasing $r$ leads to a less strong decline in the UVLF with redshift.

Note that the best fit parameter values and goodness-of-fit numbers for the alternative flattening, shallow-slope, bursty (i.e. non-unity duty cycle), high scatter, and truncation models (see \S\ref{sec:alt_models} for details) are also included in Table \ref{table:params}. 
To assess whether any improvement in $\chi^2$ is significant given the additional free parameters adopted in these models we also calculate the Akaike information criterion (AIC) differences with respect to the shallow-slope model, which has the smallest AIC \citep{Banks17}. We defer our discussion of these results to \S \ref{sec:additional}.

\section{Implications}
\label{sec:implications}

We now turn to explore the implications of the CLF fits for our understanding of the SFE versus halo mass and its redshift evolution (\S \ref{sec:sfe_results}). This is, in turn, valuable for determining the properties of the sources of reionization, the expected reionization history of the universe (\S \ref{sec:ionization_hist}), and related topics 
(\S \ref{sec:census}-\ref{sec:tau_e}).  

\subsection{Star Formation Efficiency} 
\label{sec:sfe_results}

In our model, bounds on the average UV luminosity-halo mass relationship may be directly translated into constraints on the average SFE as a function of halo mass and redshift. Specifically, rearranging Equations \ref{eq:M_acc}-\ref{eq:SFR_LUV}, substituting in the parameterization of Equation \ref{eq:L_c}, and inserting some characteristic numbers we find:
\begin{equation}
    f_\star(M,z) = 0.2 \, e^{g(\sigma, M_{\text{UV},0})} \left[\frac{2(M/M_1)^{p-\delta}}{1 + (M/M_1)^q}\right] \left(\frac{1+z}{7}\right)^{r-\eta}.
    \label{eq:f_star_scaling}
\end{equation}
Here $M_1$ is the break mass scale from Equation \ref{eq:L_c}, while $p$ and $q$
are the power-law indices which characterize the halo mass dependence on either side of the break. Recall also that $r$ 
describes the redshift evolution in $L_c(M,z)$. Furthermore, $\delta$ and $\eta$ come from the scalings of the baryonic accretion rate with mass and redshift (Equation \ref{eq:M_acc}). 
 Finally,
$e^{g(\sigma, M_{\text{UV},0})}$ is a normalization factor which depends on the scatter assumed in the CLF, $\sigma$\footnote{The mean of a lognormal is related to its median by a factor of $e^{\sigma^2/2}$}, and $M_{\text{UV},0}$ as:
\begin{equation}
    g(\sigma, M_{\text{UV},0}) = \frac{1}{2}\left[\sigma^2-0.37^2\right] + \frac{2}{5}\ln10\left[-23.79 - M_{\text{UV},0}\right].
    \label{eq:f_star_normalization}
\end{equation}
In our best fit model at $z=6$ the SFE peaks at an efficiency of
$f_\star = 21\%$ close to a mass scale of $\sim M_1$, as implied by the normalization of Equation \ref{eq:f_star_scaling}.\footnote{Note, however, that $f_\star(M,z)$ does not peak {\em exactly} at mass $M=M_1$, but the scaling of Equation \ref{eq:f_star_scaling} is nevertheless illustrative. The peak mass is, in general, a function of $p$ and $q$ but will be near $M_1$ for models similar to our B21+22 best fit.} 

\begin{figure}[ht]
  \centering
  \includegraphics[width=0.45\textwidth]{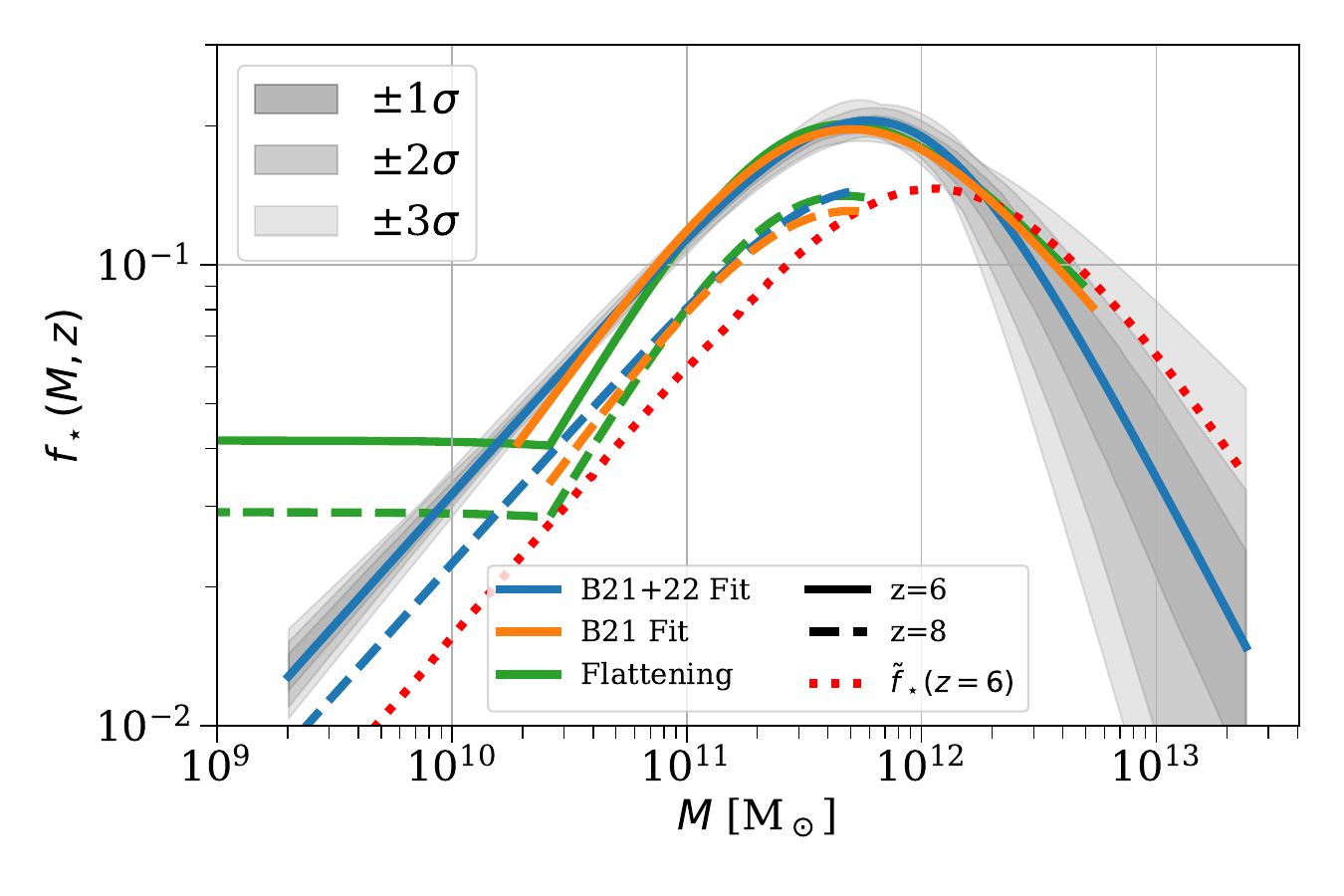}
  \caption{Constraints on the SFE versus halo mass at $z=6$ (solid) and $z=8$ (dashed) for our three models. As in the previous figure, we show only the range of halo masses probed
  by each data set.  
  In the fiducial scenario, the inferred $f_\star$ follows the form of Equation \ref{eq:f_star_scaling} where at $z=6$ it peaks at $f_\star =21\%$ and a halo mass of $M = 6\times10^{11}\text{M}_\odot$.  The shaded bands represent the effects of uncertainties in model parameters on our $z=6$, B21+22 fit at the 1-, 2-, and 3-$\sigma$ confidence levels. These constraints, while illustrative, only show the allowed range for the UV luminosity-halo mass relationship model of Equation \ref{eq:L_c} and do not capture the uncertainties in the functional form assumed here. The flattening scenario, hence, may lie outside of this band yet is also compatible with the current measurements. The red dotted line gives $\tilde{f}_\star(M,z=6)$ (see text) for comparison.
    }
  \label{fig:f_star}
\end{figure}

Figure \ref{fig:f_star} shows the average SFE fits in our fiducial models, as well as in the contrasting flattening case. In our fiducial model at $z=6$, the combined B21+22 data constrain the star formation across about four orders of magnitude in halo mass. The fits indicate
a well defined peak in SFE near a halo mass of $M \sim 6 \times 10^{11} {\rm M}_\odot$, with $f_\star(M) = 0.21$ at the peak.
The SFE declines towards small halo mass as 
$f_\star(M) \propto M^{p-\delta} \propto M^{0.56}$ and drops more steeply as
$f_\star(M) \propto M^{p-q-\delta} \propto M^{-1.02}$ at high mass. The grey bands in Figure \ref{fig:f_star} illustrate the uncertainties in the SFE for our fiducial CLF model at $z=6$. Within this class of models, the trends with halo mass are well determined, although alternative scenarios are still viable (see \S\ref{sec:alt_models}). 

For comparison we also compute $\tilde{f}_\star(M) = M_\star/(M \, \Omega_b/\Omega_m)$, the stellar mass-baryonic mass ratio,
as often considered in the literature and sometimes also referred to as the ``star formation efficiency'' although this is of course distinct from our definition here (see also \citealt{Furlanetto2017}).
This calculation uses Equation \ref{eq:M_acc} to determine the mass history of each halo at earlier times, and we combine this with our knowledge of how star formation rate/efficiency evolves with halo mass and redshift (Equations \ref{eq:sfr_efficiency} \& \ref{eq:f_star_scaling}) to compute the corresponding stellar mass.
The resulting $\tilde{f}_\star(M)$ at $z=6$ is shown as the red dotted curve in Figure \ref{fig:f_star}. It is a bit smaller in normalization than $f_\star$, at least below the peak mass. This is because the stellar mass reflects the cumulative history of the star formation in each halo and the SFE is lower at smaller masses and at earlier times. Nevertheless, the low mass power-law index for
$\tilde{f}_\star(M)$ and $f_\star(M)$ are nearly identical. 

Our results regarding the mass dependence of $f_\star$ and $\tilde{f}_\star$ are suggestive of the feedback-regulated star formation picture, as often invoked at lower redshift (e.g. \citealt{Behroozi:2014tna}, as well as in previous studies at high redshift, e.g. \citealt{Sun2016,Furlanetto2017}). In this case, the declining $f_\star(M,z)$ at small masses may reflect the combined impact of supernova and photoionization feedback. In fact, our best fit scaling is very similar to that found in the
FIRE zoom-in hydrodynamic simulations of galaxy formation \citep{Ma:2017avo}. These authors measure the stellar to total baryonic mass as a function of dark matter halo mass, averaged over their simulation outputs at $z \sim 5-12$ finding
$\tilde{f}_\star \propto M^{0.58}$. In the FIRE simulations, supernova feedback regulates star formation and leads to a power-law index which lies within
the $1-\sigma$ allowed range from our fiducial CLF results.\footnote{At masses much below the peak in SFE, the trend of $\tilde{f}_\star(M)$ with mass in our models matches that of $f_\star(M)$ and so this is a fair comparison.} 
However, the average normalization of $\tilde{f}_\star$ in the FIRE galaxies is a factor of roughly $\sim 2$ smaller than in our fits at $z \sim 6$. This difference could arise because the FIRE results are averaged over a broad redshift range, but their SFE appears to evolve little over the relevant redshifts. Alternatively, our conversion between UV luminosity and SFR might be inaccurate (Equation \ref{eq:SFR_LUV}), or this could point to deficiencies in the sub-grid feedback models adopted in FIRE.  In any case, the match between our inferred scaling and the predictions from galaxy formation simulations is suggestive and consistent with the notion that supernova feedback regulates star formation in small mass halos. 

At large mass scales, $M \gtrsim M_1$, our $z=6$ results also show evidence for a decline in SFE. The $z=8$ results, however, only reach slightly beyond $M_1$ and do not show strong evidence for a decline. This may relate to the fact that the UVLF measurements at $z \geq 8$ do not yet sample the bright end of the luminosity function.\footnote{Note that at $z=6$ the dust correction (Equation \ref{eq:dust_A}) at the bright end is about a factor of ten in luminosity, while a {\em stronger} correction would be needed to remove the evidence for a declining SFE towards high mass. Hence dust uncertainties are unlikely to result in a spurious decline in SFE at high mass.} At lower redshifts, where a similar decline in SFE at high mass is found (e.g. \citealt{Behroozi:2014tna}), it is common to ascribe the drop in SFE at high masses to AGN feedback. An interesting question is whether AGN can be responsible for this effect even at $z \sim 6$. Although previous analyses indicated a sharp decline in the AGN luminosity function at high redshift \citep{Kulkarni:2018ebj}, recent JWST studies suggest the existence of additional populations of faint AGN at $z>6$ \citep{harikane2023jwstnirspec,fujimoto2023uncover}. Nevertheless, further work is required to explore whether the declining SFE at these redshifts owes to AGN feedback. Note that while the FIRE simulations do not show evidence for a high mass decline in $f_\star(M,z)$ \citep{Ma:2017avo}, these simulations are incomplete at high masses and do not include AGN feedback. 

In our best fit fiducial model the normalization of the SFE declines towards higher redshifts. As mentioned earlier, the redshift dependence follows
$f_\star(M,z) \propto (1+z)^{r -\eta}$ (Equation \ref{eq:f_star_scaling}), reflecting evolution in both the UV luminosity-halo mass relation and the matter accretion rate. In our best fit model $r=1.12$, but $\eta=2.5$ \citep{McBride2009} (see also Equation \ref{eq:M_acc}), and so even though the UV luminosity at fixed halo mass increases with redshift this is outpaced by the growing mass accretion rate. The physical origin of our derived SFE evolution is far from obvious and is worthy of future study. Our fits differ from, for example, the FIRE simulations where the SFE appears fairly constant with redshift \citep{Ma:2017avo}. One possibility is that our model may partly mis-attribute evolution in the relationship between UV luminosity and SFR (Equation \ref{eq:SFR_LUV}) to evolution in the SFE. 

Also, a potential limitation of the assumed functional form in our fiducial model (Equation \ref{eq:f_star_scaling}) is that it assumes a redshift invariant double power-law mass dependence. 
 Note that it is only at the low redshift end of our sample that current data really probe the full {\em shape} of $f_\star(M,z)$ in the model. This is because probing the turnover mass scale requires sampling the bright end of the UVLF.
 Therefore, provided that the faint-end slope parameter $p$ also evolves little across the redshift range probed, our description should be a good one. 
 Although the redshift invariant form provides an adequate fit to current data, it will be interesting to consider more flexible models
in the future. 

In order to explore some possible
departures from the double power-law form, we
consider various alternatives such as the flattening case in more detail in \S \ref{sec:alt_models}. As a first illustration here, however, the green curves in Figure \ref{fig:f_star} show best fit
flattening model results. Although the $z \sim 6$ flattening model lies outside the grey band, it still provides a good fit to the data. That is, the grey band describes only the ($\pm 3-\sigma$) uncertainty {\em within the context of the fiducial double power-law model and does not necessarily disfavor other possibilities.} 

\subsection{Ionization History}
\label{sec:ionization_hist}

Our inferences of the SFE as a function of halo mass and redshift can be used to predict the reionization history of the universe, as will be discussed below. These predictions can be compared with current and future measurements. 

Although numerous studies have investigated related questions (e.g. \citealt{Robertson:2015uda,Mashian2016,Sun2016,Yung2020}), we consider the recent UVLF data from B22 which probe the role of individually dim galaxies in reionizing the universe. 

The evolution of the average ionization fraction in the IGM, $\langle x_i(z)\rangle$, is controlled by the difference between the rate of production of ionizing photons per hydrogen atom, $\dot n_{\rm ion}$, and the rate of recombinations, here characterized by a recombination time, $t_{\rm rec}$, as: \citep{Shapiro:1987zz,Madau:1998cd}:
\begin{equation}
\frac{\dif \langle x_i(z)\rangle}{\dif t} = \dot{n}_{\rm ion}(z) - \frac{\langle x_i(z)\rangle}{t_{\rm rec}(z)}.
\label{eq:xi_z}
\end{equation}
In this context, note that $\dot n_{\rm ion}$ receives contributions from all of the ionizing sources, including any sources of radiation that are too faint to detect individually, even in the deep lensed UVLF samples of B22. 

In what follows we neglect the effect of alternative sources of ionizing photons, such as AGN, but do extrapolate our models to cover star formation in all halos above a minimum mass $M_{\rm min}$. We assume that $\dot n_{\rm ion}$ is proportional to the star formation rate density, i.e., to the average star formation rate per unit co-moving volume. 
This quantity, denoted here by $\rho_{\rm SFR}$, is set by the average SFR-halo mass relationship in our models as:
\begin{equation}
    \rho_{\rm SFR}(z) = \int_{M_{\rm min}}^\infty \textrm{SFR}(M,z) n(M,z)\dif M.
    \label{eq:rho_SFR}
\end{equation}
As usual, the SFR follows directly from the SFE, as determined by our CLF fits, along with the matter accretion rate (see Equations \ref{eq:M_acc}-\ref{eq:sfr_efficiency}).

In order to predict the ionizing photon production rate per hydrogen atom (reaching the IGM), we need to further specify the typical ionizing spectrum of each galaxy, and the escape fraction of ionizing photons. Here we parameterize the spectrum through $N_\gamma$, the number of ionizing photons produced per stellar baryon, which then connects $\dot n_{\rm ion}$ and $\rho_{\rm SFR}$.\footnote{Note, however, that for this calculation it is unnecessary to go through the intermediate step of modeling the star formation rate density. Instead, we could have directly predicted the ionizing photon production rate from the UV luminosity density.
We nevertheless phrase our results in terms of $\rho_{\rm SFR}$, since this is a relatively familiar and intuitive quantity. Put differently, here $\dot n_{\rm ion}$ only depends on the product of $\kappa_{\rm UV}$ and $N_\gamma$ rather than on each of these individually.}
As in \cite{Sun2016} we adopt $N_\gamma = 4,000$. The escape fraction of ionizing photons, $f_{\rm esc}$, accounts for the fact that many of the ionizing photons produced are absorbed within galaxies and only a fraction of such photons escape to ionize atoms within the IGM. Unfortunately, the escape fraction during reionization is empirically and theoretically uncertain but we adopt $f_{\rm esc}\sim0.1-0.2$ as our baseline range for comparison, motivated by previous studies \citep{robertson_etal2013,Ishigaki18}. We also assume that $N_\gamma$ and $f_{\rm esc}$ are independent of halo mass and redshift.

Inserting some characteristic numbers, the ionizing photon production rate per hydrogen atom is: 
\begin{equation}
\begin{split}
\dot{n}_{\rm ion}(z) = & 5.73 \text{ Gyr}^{-1} \left[\frac{A_{\rm He}}{1.22}\right] \left[\frac{N_\gamma}{4000}\right] \left[\frac{f_{\rm esc}}{0.2}\right] \\ & \times \, \left[\frac{\rho_{\rm SFR}(z)}{3.7\times10^{-2}\text{M}_\odot\text{ Mpc}^{-3}\text{ yr}^{-1}}\right] \left[\frac{\Omega_b}{0.049}\right]^{-1}. \label{eq:n_ion}
\end{split}
\end{equation}
Here $A_{\rm He}=\frac{4}{4-3Y_p}$, with $Y_p$ the primordial mass fraction of helium, is a correction factor due to helium (e.g. \citep{Sun2016}), $\Omega_b$ is the baryon density parameter, and $\rho_{\rm SFR}(z=6) = 3.7\times10^{-2} \, \text{M}_\odot\text{ Mpc}^{-3}\text{ yr}^{-1}$ in our best fit model (to the B21+B22 data).
Since the age of the universe at $z \sim 6$ is comparable to 1 Gyr,
the characteristic rate at $z=6$ above corresponds to several photons per hydrogen atom over the age of the universe.

Recombinations have a sub-dominant effect on the average ionization history: the second term on the right-hand side of Equation \ref{eq:xi_z} is typically 20-30\% as large as the ionization term during most of reionization. More specifically, the average time between recombinations is:
\begin{equation}
    \begin{split}
    t_{\rm rec} =& 0.62\text{ Gyr} \left[\frac{C_{\text{H}_{\rm II}}}{3}\right]^{-1}\left[\frac{T_e}{10^4\text{ K}}\right]^{0.76} \\ &\times \left[\frac{\Omega_b}{0.049}\right]^{-1}\left[\frac{1+z}{7}\right]^{-3}.
    \end{split}
\end{equation}
Here $T_e$ is the temperature of the IGM (near the cosmic mean density), the scaling arises from the atomic physics of hydrogen recombination, and $C_{\text{H}_{\rm II}}$ is the clumping factor of ionized hydrogen in the IGM, defined as $C_{\text{H}_{\rm II}} \equiv\frac{\langle \rho^2_{\text{H}_{\rm II}}\rangle}{\langle \rho_{\text{H}_{\rm II}}\rangle^2}$. We adopt
redshift independent values of
$C_{\text{H}_{\rm II}}=3$ and $T_e=10^4$K (see e.g. \citealt{Sun2016} and references therein for a discussion). 

\begin{figure}[ht]
  \centering
  \includegraphics[width=0.45\textwidth]{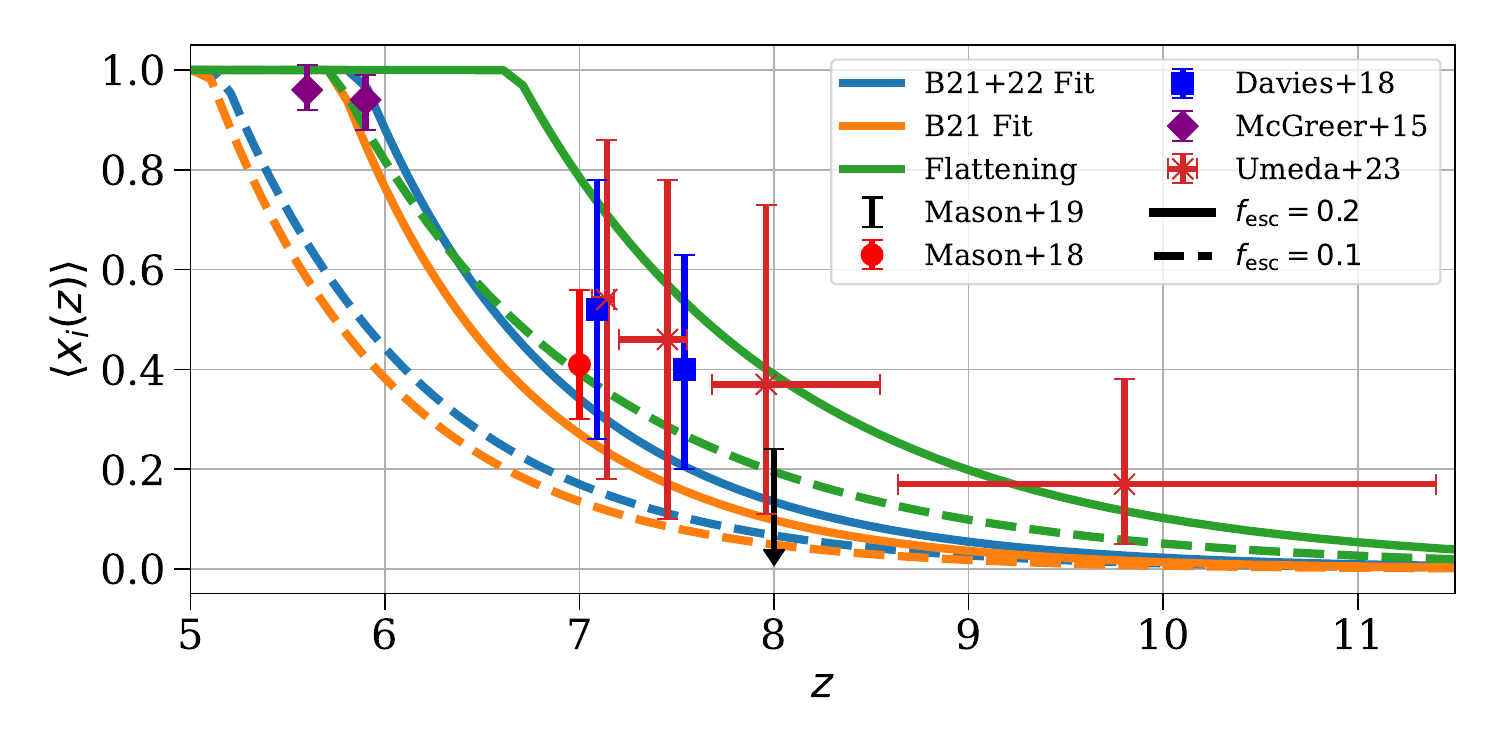}
  \caption{The reionization history from our best fit models, as compared to current measurements. The curves show the average ionization fraction as a function of redshift, and follow the usual color code of the previous figures. The solid curves
  assume that the escape fraction of ionizing photons is $f_{\rm esc} = 0.2$ while the dashed curves adopt $f_{\rm esc} = 0.1$.
   Reionization completes slightly earlier (i.e. at higher $z$) after including the B22 data since the faint-end luminosity function is marginally steeper in these observations, while reionization occurs at still higher redshifts in the flattening model. The data points show recent measurements from \cite{McGreer:2014qwa,Davies:2018yfp,Mason:2017eqr}, \cite{Mason:2019ixe} and \cite{Umeda23}.}
  \label{fig:xi}
\end{figure}

Figure \ref{fig:xi} shows the resulting ionization history implied by the best fit parameters in each of our three models, under the usual color code and with solid and dashed curves corresponding to $f_{\rm esc}=0.2$ and $f_{\rm esc}=0.1$ respectively. These models are compared to current ionization fraction measurements/bounds
from: dark segments in the Ly-$\alpha$ and Ly-$\beta$
 forests \citep{McGreer:2014qwa}, Ly-$\alpha$ damping-wing observations towards
 high redshift quasars \citep{Davies:2018yfp}, observations of Ly-$\alpha$ emission lines towards Lyman-break galaxies \citep{Mason:2017eqr,Mason:2019ixe}, and recent damping-wing measurements towards a stacked sample of UV luminous galaxies from JWST \citep{Umeda23}.
 We find better agreement in the B21 and B21+B22 fits with $f_{\rm esc}=0.2$, while $f_{\rm esc}=0.1$ is slightly preferred in the flattening case. 
 Assuming these values for the escape fraction, in broad agreement with previous work (e.g. \citealt{Sun2016}, \citealt{Robertson:2015uda}), we find consistency with current reionization history measurements, although 
the measurement errors at still too large to provide a strong test. 
In detail, our fiducial best fit to the lensed UVLF prefers a slightly earlier completion to reionization (assuming a fixed escape fraction) than in the fit to B21 alone. This is driven by the preference for a slightly steeper faint end UVLF after including the B22 data (\S \ref{sec:clf_fits}). In the flattening case, a still slightly earlier completion to reionization is achieved. This results because of the more abundant faint galaxy populations in this scenario (see Figure \ref{fig:LF} and \S \ref{sec:additional}).  
Hence, in addition to the large uncertainties in the escape fraction, even after including the lensed UVLF data, the importance of low luminosity galaxies remains somewhat unclear. 

\subsection{The Census of Ionizing Photons}
\label{sec:census}

\begin{figure}
    \includegraphics[width=0.23\textwidth]{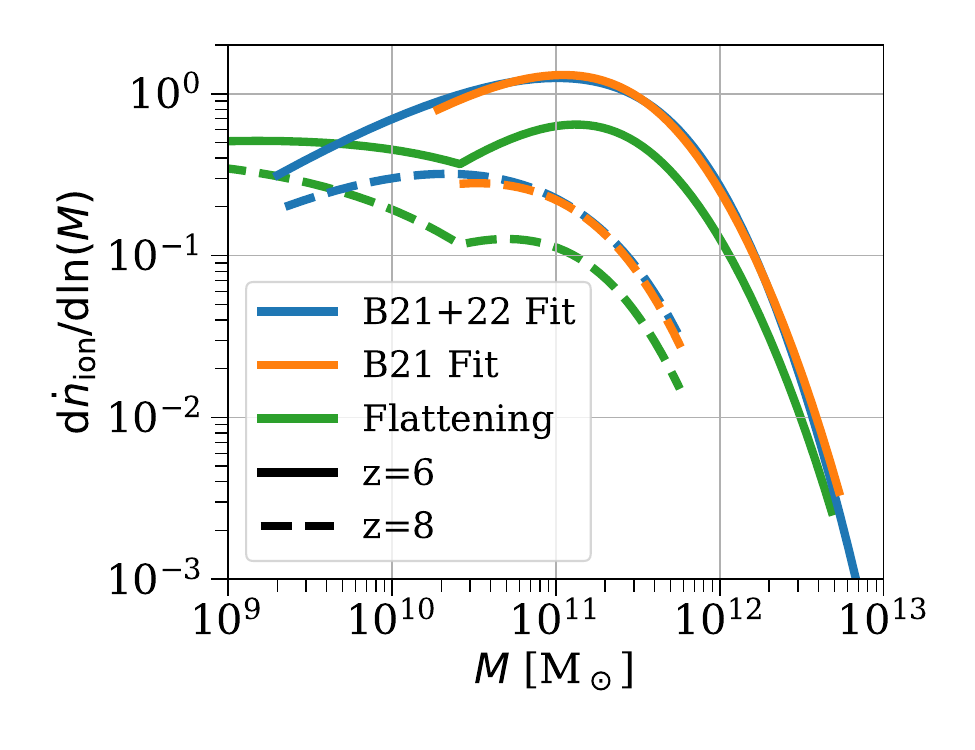}
    \includegraphics[width=0.23\textwidth]{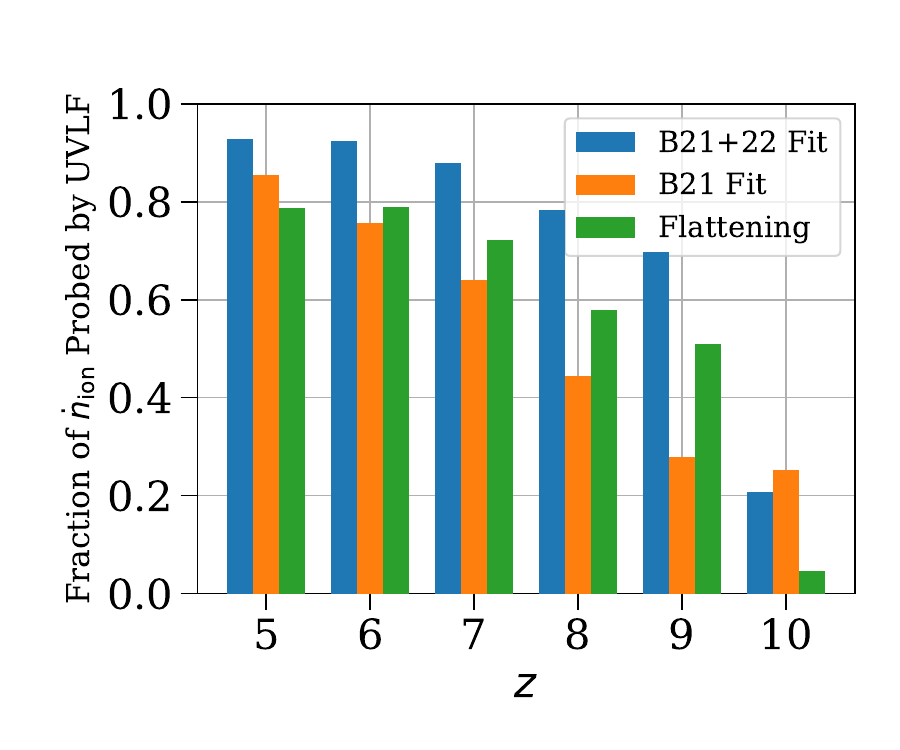}
  \caption{Census of ionizing photons from current UVLF measurements. {\em Left}: The contribution to the ionizing photon product rate budget per logarithmic interval in halo mass, ${\rm d} \dot n/{\rm d}\ln M$ vs $M$. The color codes/line-styles for the curves are the same as in previous figures. 
  The escape fractions are assumed to be independent of halo mass and set to $f_{\rm esc} =0.2, 0.2, 0.1$ for B21+B22, B21, and the flattening model, respectively. In the fiducial CLF models, the peak contribution at $z=6$ arises from  halos with mass slightly larger than $M \sim 10^{11} {\rm M}_\odot$ and
  the peak mass scale shifts towards lower mass at increasing redshift. In the flattening model, lower mass halos also make an important (at $z=6$) or even dominant contribution (e.g. in the $z=8$ example). {\em Right}: The fraction of the ionizing photon rate production accounted for by current UVLF measurements as a function of redshift (assuming $M_{\rm min} = 10^8 {\rm M}_\odot$). In our fiducial model, the lensed UVLF measurements account for $\geq 80\%$ of the ionizing photons at $z \lesssim 8$. The field only (B21) observations are less complete, while the contributions from low luminosity galaxies below the reach of current surveys are more important in the flattening model. The untouched region beyond $z \gtrsim 10$ is being probed by JWST. 
    }
    \label{fig:n_ion}
\end{figure}

We can further address the extent to which the HFF UVLF measurements, and those in the field, capture the galaxies responsible for reionizing the universe. Do the abundant but individually dim sources largely drive reionization or are the more luminous yet rarer sources most important? 
As already alluded to, this answer depends somewhat on the model 
assumed.

To approach this question, in the left panel of Figure \ref{fig:n_ion} we consider the relative contributions to $\dot n_{\rm ion}$ from different (logarithmic) mass bins. This characterizes the relative importance of
galaxies in reionizing the universe as a function of their host halo mass (see also \citealt{Yung2020}). Since
\begin{equation}
    \frac{\dif \dot n_{\rm ion}}{\dif \ln M} \propto \frac{\dif \rho_{\rm SFR}}{\dif \ln M} =  {\rm SFR}(M) M n(M),
    \label{eq:dnion}
\end{equation}
the results will, in general, depend on the precise SFR$(M)$ relation assumed.

The figure shows that, in our fiducial CLF models, the peak in $\dif \dot n_{\rm ion}/\dif \ln M$ is at a mass slightly larger than $M \sim 10^{11} {\rm M}_\odot$ at $z \sim 6$, while it moves to a mass scale slightly smaller than $M \sim 10^{11} {\rm M}_\odot$ by $z=8$. Although smaller mass halos are more abundant, their SFE is sufficiently small so such halos play a sub-dominant role. The B21+B22 curves extend to lower halo mass than B21 alone, showing how the additional reach towards low luminosity in the HFFs helps to enumerate the contribution from galaxies in smaller mass halos.

However, as discussed further in \S \ref{sec:additional} the flattening case is still allowed by the present data. In this case, the $z \sim 8$ $\dif \dot n_{\rm ion}/\dif \ln M$ curve (dashed green in the left-hand panel of Figure \ref{fig:n_ion}) actually peaks at a halo mass beneath the reach of even the lensed UVLF measurements. Consequently, although $\dot n_{\rm ion}$ is insensitive to the precise $M_{\rm min}$ in our fiducial CLF model since low mass halos have increasingly small star formation efficiencies, the results in the flattening case become dependent on $M_{\rm min}$ (see below for numbers).

The right-hand panel of Figure \ref{fig:n_ion}
compares the fraction of the total ionizing photon production rate, at each redshift, from halos of masses accounted for by the current UVLF measurements (using Equations \ref{eq:L_c} and \ref{eq:dnion}). Remarkably, in our baseline CLF model the inclusion of the lensed UVLF measurements allows about $90\%/90\%/80\%/70\%$ of the ionizing photon budget to be accounted for at
$z=6/7/8/9$, i.e. this implies only a small fraction comes from still fainter galaxies. This is improved from the corresponding numbers from B21 alone (i.e. without the lensed measurements) of 
$80\%/60\%/40\%/30\%$ at $z=6/7/8/9$. However, even the lensed UVLF estimates are less complete in the flattening case and the precise fraction of the ionizing photon budget accounted for in this case depends on $M_{\rm min}$. Specifically, the fractions become $80\%/70\%/60\%/50\%$ at $z=6/7/8/9$ for $M_{\rm min}=10^8{\rm M}_\odot$ and $70\%/50\%/40\%/30\%$ at $z=6/7/8/9$ for $M_{\rm min}=10^7{\rm M}_\odot$. 

Moreover, it is also possible
that the escape fraction increases towards small halo mass, as might $N_\gamma$; in such cases, the galaxies in small mass halos would play a more important role than under the assumptions of Figure \ref{fig:n_ion}. It will likely require improved measurements of the reionization history, along with refined UVLF measurements, to fully assess the role of faint galaxies in reionizing the universe. Estimates
of the escape fraction of ionizing photons, and its dependence on galaxy properties, will also be important. 

On the other hand, we can expect improvements at $z \gtrsim 9$ from the JWST. Measurements of the lensed UVLF and in the field from the JWST may extend the current census of Figure \ref{fig:n_ion} out to $z \sim 12-15$ or so at a comparable level of completeness, with the precise limits dependent on the still-uncertain $z\gtrsim10$ UVLF and the volume probed in JWST lensing fields (see \S \ref{sec:jwst}). 

\subsection{Ionizing Emissivity \& The Escape Fraction}
\label{sec:emissivity}

\begin{figure}[ht]
  \centering
  \includegraphics[width=0.45\textwidth]{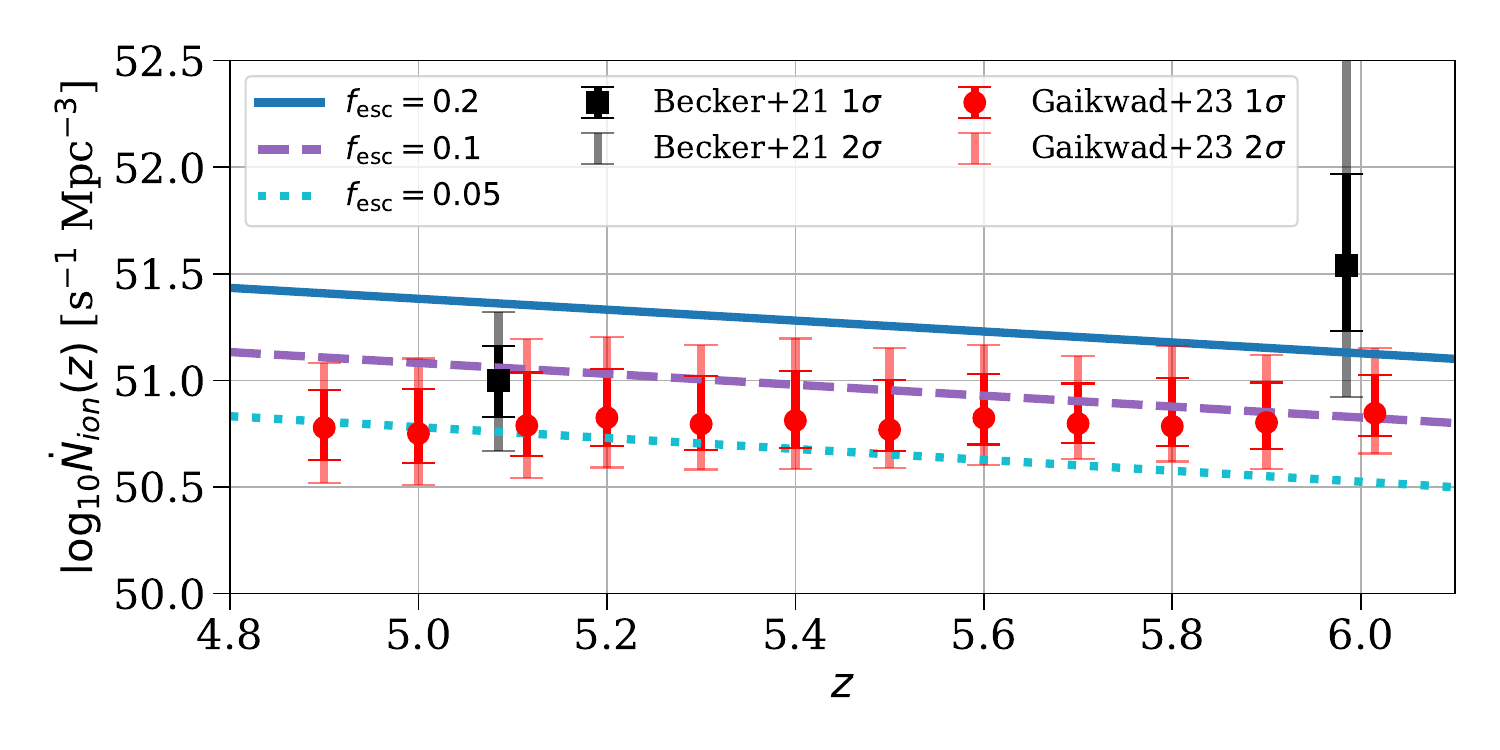}
  \caption{The ionizing emissivity as a function of redshift compared to inferences from Ly-$\alpha$ forest observations at $z \leq 6$.
   The dotted cyan, dashed purple, and solid blue curves show the emissivity predicted in our fiducial B21+B22 model fit for $f_{\rm esc} = 0.05, 0.1$, and $0.2$. The black squares with $1-\sigma$ (black) and $2-\sigma$ (grey) error bars are constraints from \cite{Becker21} and red circles with $1-\sigma$ (red) and $2-\sigma$ (pink) error bars come from \cite{Gaikwad:2023ubo}. Points at $z=5.1$ and $z=6$ that overlap have been slightly displaced in redshift for visual clarity. The \cite{Gaikwad:2023ubo} measurements prefer $f_{\rm esc} \sim 0.05-0.1$ over higher escape fractions.}
  \label{fig:emissivity}
\end{figure}

It is also instructive to model the ionizing emissivity, i.e. the rate of production of ionizing photons per unit co-moving volume as a function of redshift. At $z \lesssim 6$ this quantity can be inferred from observations of the Ly-$\alpha$ forest. More specifically, the mean transmission through the Ly-$\alpha$ forest can be estimated and this, in turn, implies constraints on the average hydrogen photoionization rate, $\Gamma_{\rm HI}$. This step generally requires comparisons with numerical simulations of the Ly-$\alpha$ forest. One further requires knowledge of the mean-free path to ionizing photons which
can also be estimated from Ly-$\alpha$ forest data. 

The ionizing emissivity inferred from this procedure at $z \sim 5, 6$ by \cite{Becker21} is
shown in Figure \ref{fig:emissivity} (see also that work for a discussion of the uncertainties in these measurements). 
The total rate of ionizing photon production in our model per unit volume may be computed as:

\begin{equation}
    \dot N_{\rm ion} = \dot n_{\rm ion} \bar{n}_b \propto f_{\rm esc} N_\gamma,
\end{equation}
where $\bar{n}_b$ is the average co-moving abundance of baryons. 
For present purposes, the most interesting aspect of comparing this model with the measurements is that it can be used to 
determine the uncertain product of $f_{\rm esc}$ and $N_\gamma$ \citep{Kuhlen12,Sun2016}. This then provides a consistency test regarding the reionization history model of the previous two sections, which required $f_{\rm esc} \sim 0.1-0.2$ (assuming $N_\gamma=4,000$).\footnote{Strictly speaking, both measurements are sensitive only to the overall product of $N_\gamma$ and $f_{\rm esc}$. Although we fix $N_\gamma=4,000$ and vary $f_{\rm esc}$ around $f_{\rm esc} \sim 0.1-0.2$, one can think of this step instead as exploring variations in the product of these quantities around $N_\gamma f_{\rm esc} = 400-800$. }

Figure \ref{fig:emissivity} shows the ionizing emissivity and its redshift evolution in our fiducial CLF fit (to B21+B22) for each of 
$f_{\rm esc}=0.2$ (solid blue), $0.1$ (dashed purple), and
$0.05$ (dotted cyan). Encouragingly, the
$f_{\rm esc} = 0.2$ model lies within the $2-\sigma$ error range from the 
\cite{Becker21} measurements at each of $z = 5$ and $z=6$. However, the recent \cite{Gaikwad:2023ubo} points prefer a smaller value of $f_{\rm esc} \sim 0.05-0.1$. 

When combined with the reionization history inferences of the previous section and the electron scattering optical depth results of the next section, this might hint that the escape fraction is evolving and is smaller by the  
redshifts probed with the Ly-$\alpha$ forest than during most of reionization. Alternatively, it could indicate an inadequacy in our modeling or systematic errors in one or more of the measurements.

Overall, the errors on the emissivity measurements are still relatively large and so we conclude that our model for the ionizing sources -- which matches the UVLF data from B21+B22 -- is still in general agreement with the $z \sim 5-6$ ionizing emissivity inferred from the Ly-$\alpha$ forest in addition to the reionization history measurements (\S\ref{sec:ionization_hist}) when we adopt a constant
$f_{\rm esc} \sim 0.1-0.2$ and $N_\gamma = 4,000$. This is similar to the conclusion drawn by \cite{Sun2016} from earlier UVLF measurements in the field.

\subsection{Thomson Scattering Optical Depth}
\label{sec:tau_e}
\begin{figure}[ht]
  \centering
  \includegraphics[width=0.45\textwidth]{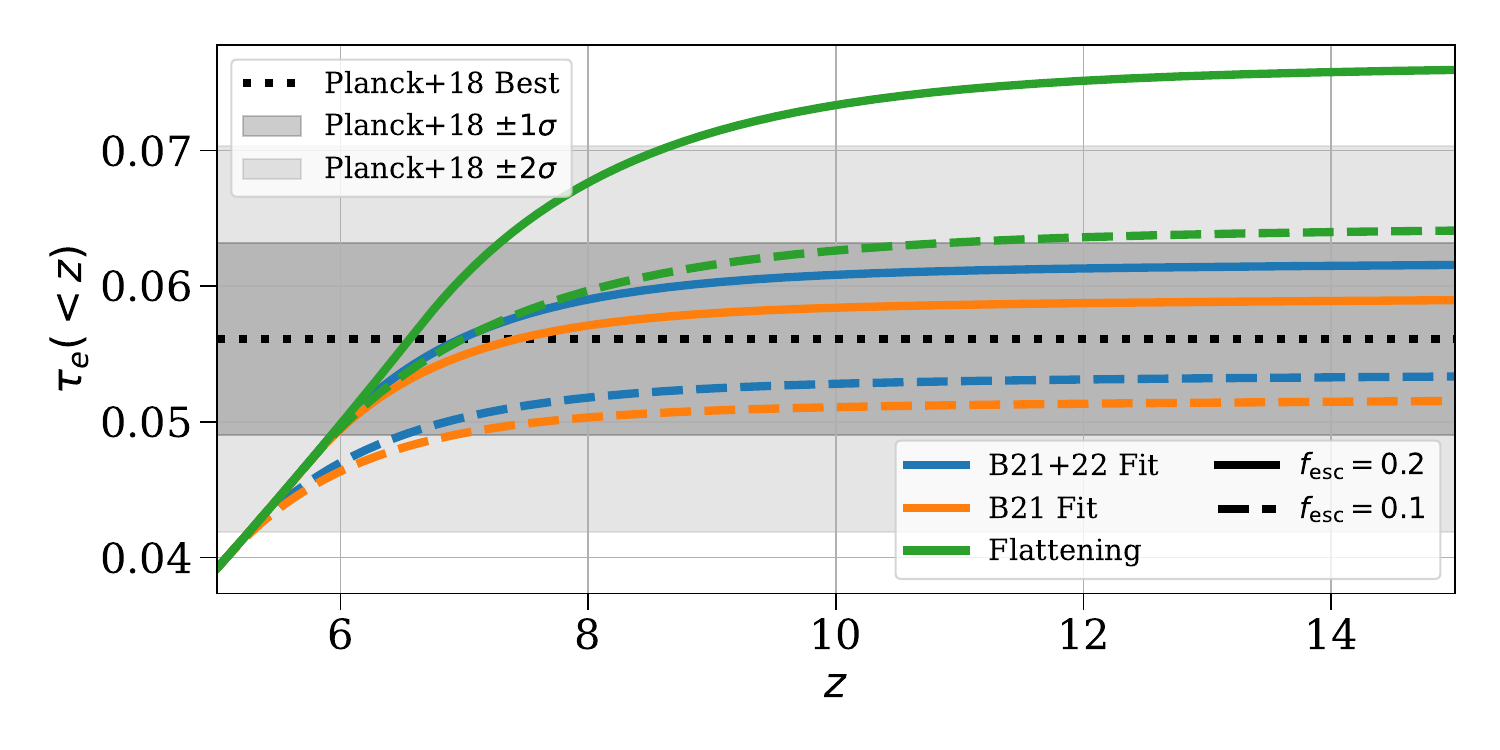}
  \caption{Electron scattering optical depths at redshifts $<z$ in our models compared to Planck 2018 $\tau_e$ measurements. Models are labeled with the usual color code, and escape fractions of $f_{\rm esc}=0.1 (0.2)$ are shown as solid (dashed) curves.  
  In the flattening model an escape fraction of $f_{\rm esc}\gtrsim 0.15$ is disfavored (at $\gtrsim 2-\sigma$ confidence), while $f_{\rm esc}\sim0.1-0.2$ are consistent with current measurements in the fiducial models.
  The best fit optical depth inferred from Planck 18 measurements is shown by the black dotted line, while the grey bands give the $1-$ and $2-\sigma$ measurement errors \citep{planck2018}.  
    }
  \label{fig:tau}
\end{figure}

A further test is provided by measurements of the optical depth of CMB photons to Thomson scattering, $\tau_e$. This provides an integral constraint on the ionization history and so bounds
the cumulative effects from all of the ionizing sources during the EoR. It complements the UVLF measurements, which are not directly sensitive to the ionization state of the IGM, and current measurements of the reionization history/ionizing emissivity (see Figure \ref{fig:xi}-\ref{fig:n_ion}) which are confined mostly to $z \leq 8$. The statistical errors on $\tau_e$ are also much smaller than the current reionization history/ionizing emissivity measurement errors.

To name one example of where the $\tau_e$ constraints are especially powerful, note that a sufficiently large optical depth could indicate an additional population of high redshift $z \gtrsim 10$ ionizing sources; the B22 UVLF and current reionization history measurements would be entirely blind to such sources.  Instead, as is widely appreciated, the Planck 2018 $\tau_e$ measurements are consistent with a relatively rapid and late completion to reionization. Exotic early source populations are not required, and are in fact bounded, by the $\tau_e$ measurements. Indeed, these measurements are entirely consistent with our baseline UVLF fits and $f_{\rm esc} \sim 0.1-0.2$ model.

More quantitatively, the average electron-scattering optical depth out to a redshift $z$ is given by:
\begin{equation}
    \tau_{e}(<z) = c \sigma_T n_H\int_0^z \frac{(1+z)^2}{H(z)} x_i(z)\left(1+N_{\rm He}(z)\right) \dif z.
    \label{eq:tau}
\end{equation}
Here $\sigma_T$ is the Thomson scattering cross section. The quantity $N_{\rm He}(z)$ accounts for helium, which we assume to be singly-ionized along with hydrogen and then doubly-ionized below $z=3$ so that:
$N_{\rm He}(z>3) = \frac{Y_p}{4(1-Y_p)}\sim0.08$ and $N_{\rm He}(z<3) = \frac{Y_p}{2(1-Y_p)}\sim 0.16$. The observable total optical depth $\tau_e$ follows from taking the large $z$ limit
of Equation \ref{eq:tau}, while the cumulative contribution out to redshift $z$, $\tau_e(<z)$,
is helpful for understanding the impact of sources at varying redshifts. 

Figure \ref{fig:tau} shows $\tau_e(<z)$ for our B21+22, B21, and flattening models for escape fractions of $f_{\rm esc}= 0.1$ and $0.2$, along with Planck 2018's best fit $\tau_e$ and the $1-\sigma$, $2-\sigma$ confidence intervals (from their TT,TE,EE+lowE+lensing+BAO results, \citealt{planck2018}). The B21+22 (B21) model gives
$\tau_e=0.062$ ($\tau_e=0.059$) for $f_{\rm esc}=0.2$, each of which lies within the $1-\sigma$ allowed region from Planck 2018, while the flattening model predicts $\tau_e=0.065$ for $f_{\rm esc} = 0.1$, just outside of the $1-\sigma$ range. Hence all three cases are consistent with the current $\tau_e$ measurements. In the flattening model, low luminosity sources beyond the reach of even B22 play an important role and this leads to a more extended reionization history (see Figure \ref{fig:xi}) and a larger $\tau_e$, although this scenario is still entirely consistent with Planck 2018 observations for $f_{\rm esc}=0.1$. The flattening case with $f_{\rm esc}=0.2$ lies just a little bit beyond the $2-\sigma$ confidence region from Planck 18.

\subsection{UV Luminosity Density \& The Detectability of the 21 cm Signal}
\label{sec:wf_coupling}

An extrapolation of our model fits to higher redshifts also has implications for the 21 cm signal from Cosmic Dawn. In particular, the 21 cm signal is expected to be observable in absorption against the CMB at early times \citep{Furlanetto:2006jb}. This is the case provided the gas kinetic temperature is cooler than the CMB temperature at the redshifts of interest, and as long as the excitation temperature (aka the ``spin temperature'') of the 21 cm line is well coupled to the gas temperature. The expectation is that UV photons from the first luminous sources will redshift into Lyman-series resonances, mix the hyperfine states, and couple the spin temperature to the gas 
temperature \citep{Wouthuysen52,Field58,pritchard_loeb2012}. 
This process is referred to as the ``Wouthuysen-Field'' (WF) effect: the upshot here is that a sufficiently intense UV radiation field is required to keep the 21 cm spin temperature from equilibrating with the CMB temperature and to yield
 a measurable 21 cm absorption signal. Therefore, we can
extrapolate the UV luminosity density in our model, calibrated to the Hubble UVLF measurements, predict the WF coupling strength, and hence the redshift onset of the 21 cm absorption signal. 

This calculation can be compared to the possible 21 cm absorption signal identified by the EDGES experiment \citep{Bowman:2018yin} (although see \citealt{Singh:2021mxo} for a recent non-detection, inconsistent with EDGES at $\sim 2-\sigma$). It also has implications for the design of future global 21 cm surveys, as well as measurements of 21 cm fluctuations. Here
we make use of the study of \cite{Madau2018}
which gives the range in UV luminosity densities
required to achieve WF coupling, as a function of the redshift at which the coupling occurs. 
We can then compare this
threshold with the extrapolated UV luminosities predicted in our various models (see also e.g \citealt{Mirocha:2018cih,Bera:2022uix,Meiksin:2023mpt,Hassan:2023asd}).  
Specifically, the UV luminosity density may be computed in our models as:

\begin{equation}
    \rho_{\rm UV} = \frac{\rho_{\rm SFR}}{\kappa_{\rm UV}} = \int_{M_{\rm min}}^\infty L(M) n(M) \dif M, \label{eq:rho_UV}
\end{equation}
while the threshold luminosity density to achieve WF coupling is approximately \citep{Madau2018}:
\begin{equation}
    \rho_{\rm UV}^{\rm WF}(z) = 10^{24.52} g_{0.06}^{-1}\left(\frac{18}{1+z}\right)^{1/2} \text{erg}\text{ s}^{-1}\text{ Mpc}^{-3}\text{ Hz}^{-1}.
    \label{eq:rho_wf}
\end{equation}

This is the UV luminosity density corresponding to a WF coupling coefficient of $x_\alpha=1$. The quantity $g$ relates to a sum over different Lyman-series resonances and accounts for redshift evolution between the emission of UV photons and scattering in one of the resonances (see \citealt{Madau2018} for details). This factor is normalized to a fiducial value of $g=0.06$.  
As in \cite{Madau2018}, we suppose that the onset of the 21 cm absorption signal corresponds to a coupling constant of $x_\alpha=1-3$ (i.e. spanning between the value
in Equation \ref{eq:rho_wf} and three times that number). We can then test at which redshifts our models (from Equation \ref{eq:rho_UV}) cross these thresholds. In comparison to the earlier calculations of \cite{Madau2018}, our analysis incorporates the more recent lensed UVLF results of B22 and also adopts a physically-motivated model to extrapolate from the UVLF measurements to higher redshifts (while \citealt{Madau2018} assumes that the UV luminosity density itself evolves as a power-law in redshift). 

\begin{figure}[ht]
  \centering
  \includegraphics[width=0.45\textwidth]{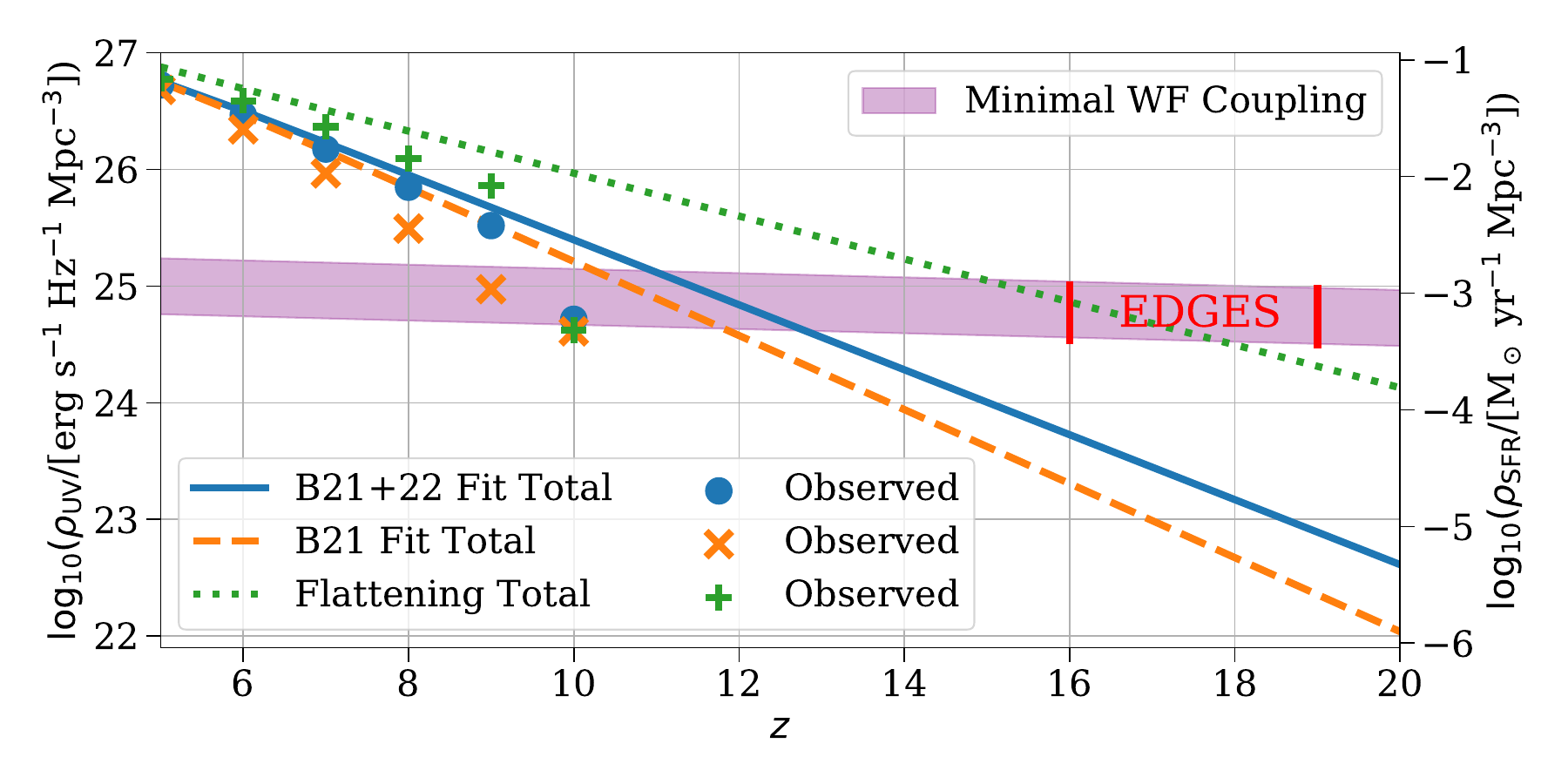}
  \caption{Implications for the onset of WF coupling between the 21 cm spin temperature and the gas temperature.
The curves give the UV luminosity density in our models, following the usual color code. The lines include extrapolations down to arbitrarily small luminosities (assuming $M_{\rm min}=10^8 {\rm M}_\odot$), while the points only consider luminosities measured in the respective data sets and redshift bins. The purple shaded band indicates the minimum range in UV luminosity densities required for UV radiation to couple the 21 cm spin temperature to the gas temperature \cite{Madau2018}. The red labeled region indicates the approximate redshift range for the onset of WF coupling implied by the EDGES measurement \citep{Bowman:2018yin}.  In the flattening model one can achieve coupling at the EDGES implied redshifts, but the onset of the 21 cm absorption signal is expected later (near $z \sim 13$) in our other models. 
}
  \label{fig:rho_UV}
\end{figure}

Figure \ref{fig:rho_UV} shows the UV luminosity density / star formation rate density in our models extrapolated to higher redshifts, as compared to the threshold densities required to achieve WF coupling. We show results for each of the CLF fits to B21+B22 (blue), B21 alone (orange), and in the flattening model fit to B21+B22 (green). For contrast, we also show (points labeled ``observed'' in legend) $\rho_{\rm UV}$ implied by the models without extrapolating to fainter luminosities than currently observed.

In our fiducial CLF models, including faint sources and extrapolating trends in redshift,
we find that coupling is achieved at
$z \sim 12-13$. In these scenarios, we expect the absorption signal to be observable only at lower redshifts than implied by EDGES. Here, the sweet spot for global 21 cm measurements
appears to be around a frequency of $\nu=1420 \, {\rm MHz}/(1+z) \sim 100-110\, {\rm MHz}$, which unfortunately lies in the FM radio band. However, one should keep in mind that the 21 cm absorption signal also depends on heating from e.g. X-rays, and so the WF coupling coefficient provides an incomplete descriptor of the signal. Interestingly, in the flattening model one expects an onset redshift close to that of the EDGES signal, and so models with rather efficient star formation in small mass halos could achieve the timing of the EDGES absorption feature (see also \citealt{Mirocha:2018cih}). Even in this case, however, the depth and the shape of the EDGES feature are surprising (e.g. \citealt{Mirocha:2018cih}). 

The symbols marked ``Observed", without extrapolation towards low halo mass, are relevant for the interpretation of current 21 cm fluctuation measurements. For example, the HERA collaboration recently used upper limits on the 21 cm fluctuation power spectrum at $z=7.9$ and $z=10.4$ to place a lower bound on the amount of X-ray heating at these redshifts \citep{HERA:2022wmy}. That is, a cold IGM might lead to larger 21 cm fluctuations than observed. This argument requires that there are significant amounts of neutral gas at these redshifts (as strongly suggested by e.g. the Planck $\tau_e$ measurements, \S \ref{sec:tau_e}), and that the 21 cm spin temperature is coupled to the gas temperature. The discrete points in the figure show that currently observable galaxies should easily yield WF coupling at $z \sim 7.9$ and that coupling at $z \sim 10.4$ requires only a modest extrapolation in redshift and luminosity beyond those of current UVLF measurements. This further validates the bounds in \cite{HERA:2022wmy}, although we caution that the {\em average} WF coupling considered here provides only a loose figure of merit.  

\section{Comparison with Early JWST Results}
\label{sec:jwst}

The JWST has already started to make revolutionary new measurements of high redshift galaxy populations \citep{Donnan22,Finkelstein_2023,Bouwens_2023,Harikane_2023}. Interestingly, these early results suggest large populations of UV luminous galaxies even at $z \sim 11-17$. 
In addition, some galaxies have estimated stellar masses as large as $M_\star \gtrsim 10^{10}-10^{11}{\rm M}_\odot$ at redshifts as high as $z \sim 9$ \citep{Labbe23}. Most of these detections are photometric candidate galaxies, selected via the Lyman-break technique, and so their redshift estimates await spectroscopic confirmation. Thus far, where spectroscopic confirmations do exist \citep{harikane2023pure} they mostly agree -- at least qualitatively -- with the findings from the photometric galaxies. Nevertheless, it will be important to further assess the purity of the $z \gtrsim 10$ JWST photometric samples: since the abundance of potential lower 
redshift interloping galaxies likely outnumbers that of the $z \gtrsim 10$ galaxies by a large factor, even interlopers with rare/unusual spectra could be an important source of contamination \citep{Furlanetto22}.
For example, a $z \sim 16$ candidate galaxy \citep{Donnan22} turned out to be a dusty $z \sim 5$ star-forming galaxy with strong H-$\alpha$ and [OIII] emission lines \citep{Arrabal23}, as suggested earlier
based on a secondary redshift solution and nearby $z \sim 5$ neighboring galaxies \citep{Naidu22} (see also \citealt{Zavala23}).  In addition, updated instrumental calibrations have led to downward revisions in the photometric redshift estimates for at least some candidate galaxies from early JWST images \citep{Adams23}. Also, the current UVLF estimates are sensitive to the precise likelihood threshold adopted for accepting candidate galaxies \citep{Bouwens_2023}.

Finally, the total number of JWST candidate galaxies and the volume probed are still relatively small. Hence, while the early JWST results may suggest more $z \gtrsim 10$ star formation than previously thought, future efforts are required to achieve fully robust UVLF measurements in this era.

\begin{figure*}[t]
  \centering
  \includegraphics[width=0.95\textwidth]{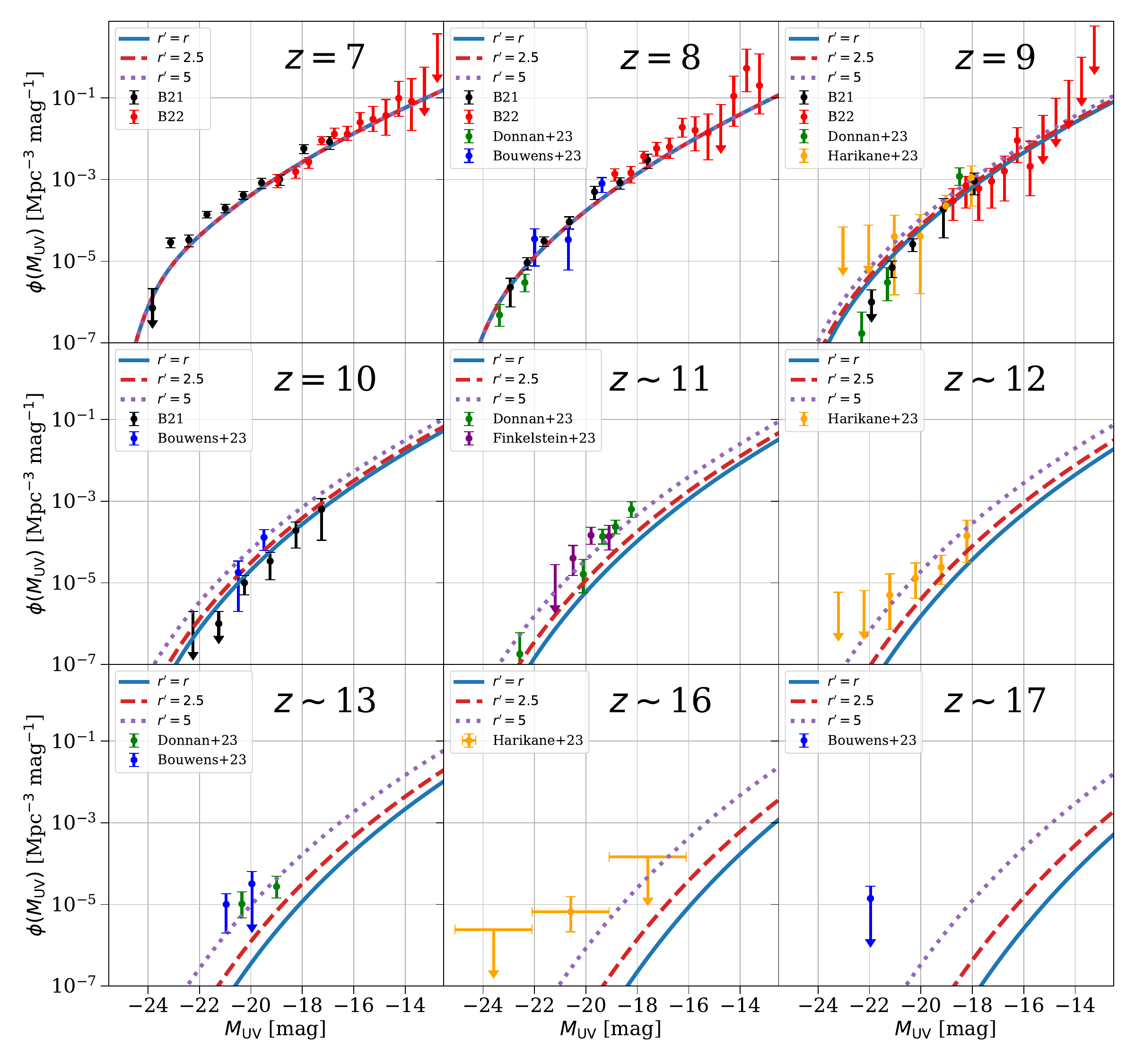}
  \caption{The redshift evolution required to explain early JWST measurements. The blue, purple, green, and gold points (and upper bounds) show preliminary JWST UVLF estimates from $z \sim 7-17$, while the black/red points show the usual B21/B22 HST points (confined to $z \leq 10$). The blue curve shows our fiducial model, while the dashed red and dotted purple curves allow an abrupt change in redshift scaling of the UV luminosity-halo mass relationship above $z > 8$. Specifically, the red curve takes the best fit redshift evolution parameter with $L_c \propto (1+z)^r$ (see Equation \ref{eq:L_c}) from $r=1.12$ to $r'=2.5$ above $z > 8$, while the purple curve assumes $r'=5$ above $z > 8$. These alternative scenarios would imply very high star formation efficiencies in the galaxy-hosting halos at high redshifts (see text) and require rapid evolution just beyond the redshifts previously probed by the HST observations.}  
  \label{fig:jwst}
\end{figure*}

Nevertheless, we can place the early JWST UVLF estimates in context by comparing them with our models calibrated to the latest HST data from B21+B22. Here, we find relatively good agreement between the HST UVLF estimates and the new JWST measurements at $z \lesssim 10$. However, the JWST UVLF estimates at higher redshifts land significantly above a simple extrapolation, to $z \gtrsim 10$, of our HST-calibrated best fitting model. One possibility is that the SFE stays constant or starts to increase beyond the redshifts well-probed by HST, even though our best fit model has the SFE {\em decreasing} across the redshift range we fit to (from $z = 5 -10$). To explore this, we suppose that the redshift evolution parameter transitions abruptly from the value
$r$ (see Equation \ref{eq:L_c}) to an alternate value, $r'$ at
$z > 8$ only. That is, we suppose the median UV luminosity scales as, for a given increased parameter $r'$, $L_{c,r'}(M,z) = L_c(M,z) \times \left(\frac{1+z}{9}\right)^{r'-r}$ at $z > 8$. Note that our fiducial best fit model has $r=1.12$ (Table \ref{table:params}).

Specifically, Figure \ref{fig:jwst} compares the HST points and the new JWST UVLF measurements from \cite{Donnan22,Harikane_2023,Bouwens_2023,Finkelstein_2023} for models with $r'=r$ (our fiducial model), $r'=2.5$, and $r'=5$
at $z=7-17$. As alluded to above, the new JWST measurements are consistent with the fiducial model and the HST points at $z \lesssim 10$, but this model significantly underpredicts the data at higher redshifts. In the alternative $r'=2.5$ model, the SFE becomes redshift independent at $z > 8$ (recall that $f_\star(M,z) \propto (1+z)^{r'-2.5}$ for our model, Equation \ref{eq:f_star_scaling}), yet this model still falls quite a bit below the JWST data at $z \gtrsim 13$. In order to better reconcile with the JWST case, a rather extreme scenario with $r'=5$ is required. In this case, the implied star formation efficiencies become quite high: for example, in galaxies of absolute magnitude $M_{\rm UV} = -20$, the $r'=5$ scenario implies $f_\star = 13 \%$ at $z=16$, compared to $f_\star = 3\%$ in our fiducial scenario at the same redshift and mass scale. Additionally, if the shape of $f_\star(M,z)$ is redshift invariant as in our fiducial model, the SFE at $M_{\rm UV}=-24$ for $r'=5$ is
$f_\star = 44\%$ at $z=16$. Further, the $r'=5$ case implies an abrupt reversal in the redshift evolution of the SFE to match both this behavior and the trend suggested by the HST data.

Although the early JWST results at $z \gtrsim 10$ require a rapidly increasing SFE towards high redshift, note that such scenarios do not generally overproduce the electron scattering optical depth constraints from Planck \citep{planck2018}. Specifically, the $r'=5$ scenario is consistent at the $1.4-\sigma$ level, assuming an escape fraction of $f_{\rm esc}=0.2$. The $\tau_e$ boost is only small as star-forming halos are still relatively rare at these redshifts. A larger change in $\tau_e$ would require high $r'$ {\em and} a smaller minimum host halo mass than in our fiducial scenario (which adopts $M_{\rm min} = 10^8 {\rm M}_\odot$).

This is in agreement with a range of previous studies, such as \cite{Mason:2023,Bouwens_2023,Mirocha2023}, which also found that HST-calibrated models under-predict the JWST UVLF measurements. As above, a number of these works noted that an increasing SFE towards early times could help reconcile with the JWST measurements \citep{Mason:2023,Bouwens_2023,Mirocha2023,Inayoshi_2022,harikane2023pure}. Although the physical origin of this possibly enhanced SFE remains unclear, \cite{dekel2023efficient} propose
that feedback-free starbursts may naturally occur at the relevant redshifts and halo mass scales. 

Another potential explanation considered is the presence of a ``top-heavy" IMF at these high redshifts, where the formation of low-mass stars is suppressed by a higher gas temperature \citep{Inayoshi_2022,Mason:2023,Harikane_2023,Finkelstein_2023}. The higher CMB temperature and the less efficient cooling of the lower metallicity gas at early times might both contribute to this effect \citep{Chon22}. In this case, the conversion factor between UV luminosity and SFR, $\kappa_{\rm UV}$ (Equation \ref{eq:SFR_LUV}), would be smaller in the early universe and galaxies would be more luminous at similar star formation efficiencies. Other possibilities include an incomplete understanding of dust attenuation \citep{Mason:2023,ferrara2022stunning,Finkelstein_2023}, larger variance in the star formation rates in small halos \citep{Mirocha2023,Mason:2023,yajima2022forever22}, a greater prevalence of AGN than expected \citep{harikane2023pure}, or more exotic options including enhanced structure formation from primordial black holes or axion dark matter mini-clusters \citep{Liu_2022,Hutsi23}.
A closely related problem is that the stellar mass estimates for some of the JWST galaxies exceed expectations informed by structure formation in LCDM with \cite{planck2018} parameters \citep{Labbe23,boylankolchin2023stress,steinhardt2022templates}. 
Among other possibilities, this could be related to systematic errors in the stellar mass estimates, or it might provide another indication of efficient early star formation and/or a top-heavy IMF. 

Further work is required to solidify the early JWST estimates, especially: obtaining additional spectroscopic confirmations of galaxy candidates at $z\gtrsim10$ \citep{harikane2023pure,Curtis-Lake2023,bunker2023jades,fujimoto2023ceers}, tests of sample purity, and improved statistics. Our HST-calibrated models will continue to provide useful baseline model expectations, including for upcoming lensed measurements with JWST which should access the faint-end of the UVLF at $z \gtrsim 10$. 

\section{Alternate Models for the Faint-End Behavior and Systematic Error Tests} \label{sec:alt_models} As discussed in \S\ref{sec:Method} our parameterization of $L_c(M,z)$ as a double power-law in halo mass is motivated by previous treatments \citep{Bouwens2015,Schive:2015kza} which are in turn informed by stellar feedback effects that are broadly borne out by simulations \citep{Ma:2017avo}.
Yet, our relatively limited observational handle on low luminosity galaxies at high redshifts leaves open the possibility of alternative behavior in this regime. 
Furthermore, bursty star formation may actually lead to a higher SFE in small mass halos than otherwise expected \citep{Furlanetto22_bursty}\footnote{Note, however, that this enhanced efficiency is not observed in the FIRE simulations \citep{Ma:2017avo} even though star formation is bursty in these simulations.}, and this will likely be accompanied by an enhanced scatter. 
In the following subsections we consider cases where the star formation is enhanced or suppressed, or the scatter around the UV luminosity-halo mass relationship is varied, in small mass dark matter halos relative to our fiducial model. We also quantify the impact of 
systematic uncertainties and cosmic variance on the lensed 
UVLF measurements. 
\subsection{Enhancing Star Formation at Low Mass}
\label{sec:enhance}

First, we consider four model extensions in which the SFE is enhanced below some critical mass, $M_2$. In the first case, the SFE flattens. Second, we allow a free spectral index between
$M_{\rm min}$ and $M_2$: we find that this prefers a shallower slope while a steeper dependence is disfavored.  In the next two variants, the scatter in the UV luminosity-halo mass relation is boosted; this also increases the SFEs implied by our fits. In the first enhanced scatter model, we simply increase the variance of the CLF (Equation \ref{eq:CLF}), while in the second case we allow a duty cycle to account for periods of bursty star formation.\subsubsection{Flattening and Shallow-Slope Scenarios}

In the flattening and shallow-slope models, we suppose that the median UV luminosity-halo mass relationship scales as
$L_c(M,z) \rightarrow L_c(M,z)\times(M/M_2)^{\delta+s-p}$ between $M_{\rm min}\leq M\leq M_2$, where the new parameters $s$ and $M_2$ help quantify the departure from pure power-law behavior at small masses. Here $s$ is defined such that the SFE scales
as $f_\star(M) \propto M^s$ at low masses (see Equation \ref{eq:f_star_scaling}); in our fiducial model, the best fit result is $f_\star(M) \propto M^{0.56}$.
Alongside the five fiducial CLF parameters we can fit for $M_2$ and fix $s=0$ (allow $s$ to vary) in our flattening (shallow-slope) models. Note that our prior on $s$ in the latter case is [-0.5, 1.5], which also allows for the possibility of a {\em steepening} $f_\star(M)$ relation ($s>p-\delta$). A steeper relation might occur if feedback effects give a stronger suppression than in our baseline model (see also \S \ref{sec:trunc}): however, we find that the data disfavor this possibility.

The results of these fits are found in Table \ref{table:params}, with $f_\star(M)\propto M^{0.27}$ at small mass preferred in the shallow-slope case (the $2-\sigma$ confidence interval is $0.14<s<0.40$ and disfavors a steepening).
As detailed in the table, the other parameters vary somewhat to compensate for the changes in the model UVLFs. Notably, in both cases, the alternate behavior occurs at $<M_2 \sim 3 \times 10^{10} {\rm M}_\odot$: these are mass scales almost exclusively probed by the lensed UVLF measurements. Interestingly, these scenarios are, in fact, preferred (even accounting for a penalty for the greater model complexity) over our fiducial models with $\Delta$AICs of $12.1$ and $0$ for the flattening and shallow-slope cases compared to the fiducial value of $\Delta$AIC $=24.3$. That is, these cases are relatively favored at $3.5-\sigma$ and $4.9-\sigma$ respectively. This evidence, is however, subject to systematic uncertainties in the lensed UVLF measurements (\S \ref{sec:additional}). 

\begin{figure}[ht]
  \centering
    \includegraphics[width=0.45\textwidth]{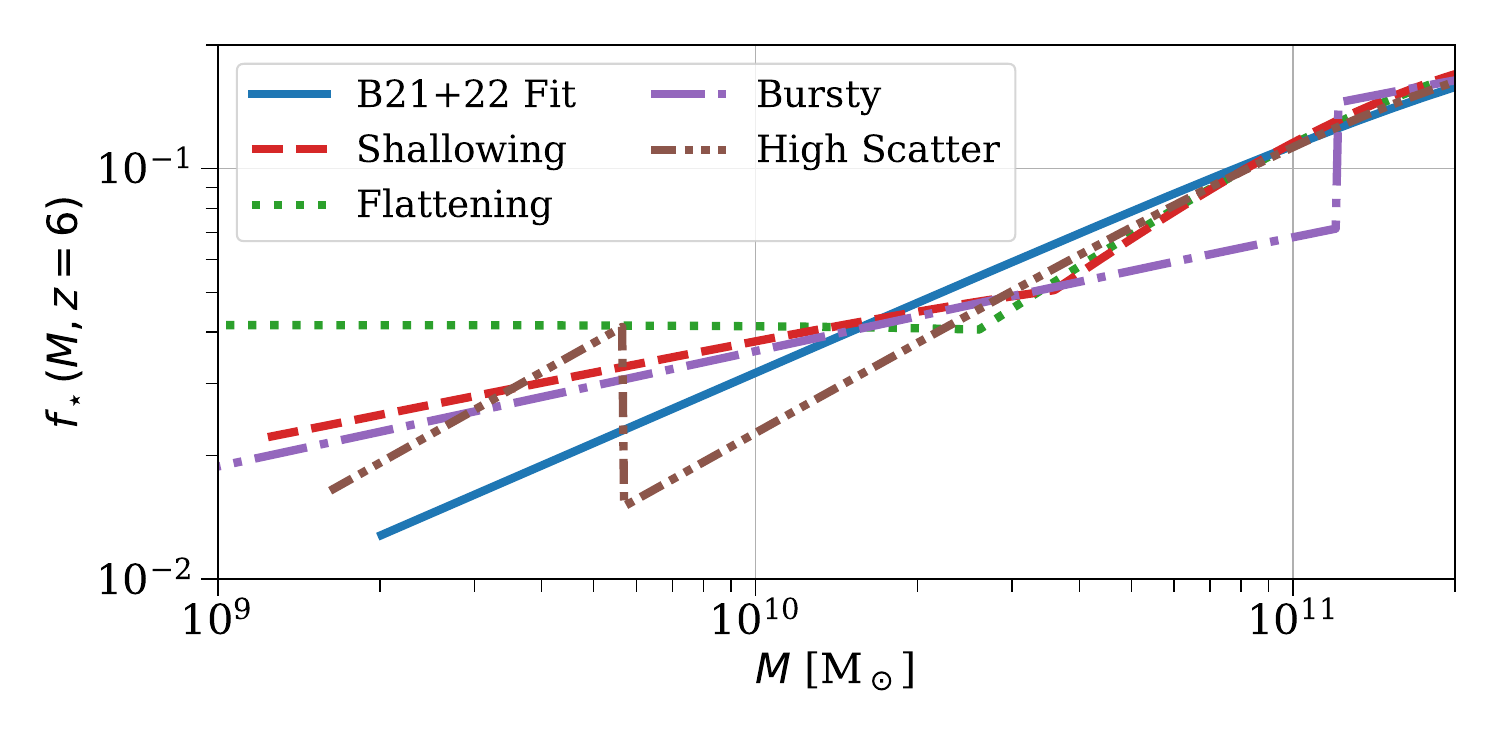}
    \includegraphics[width=0.45\textwidth]{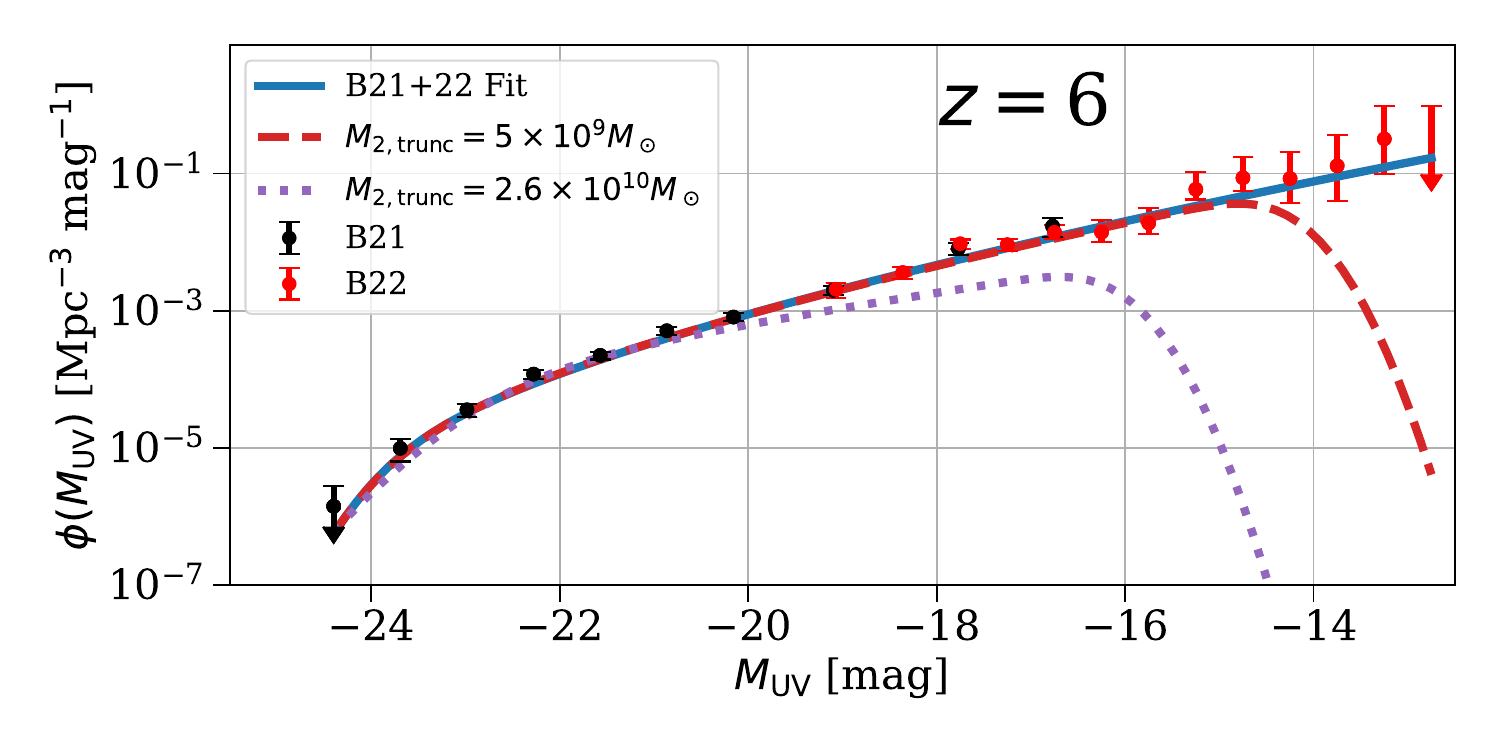}
  \caption{Effect of alternative models on the SFE and faint-end UVLF. {\em Top:} The SFE at redshift $z=6$ in our four cases of enhanced faint-end star formation compared with our fiducial best fit model (solid blue). The green dotted curve shows our usual flattening model, the red dashed case shows the best fit model with a shallower slope below $M_2$, the purple dot-dashed curve allows a non-unity duty cycle at small masses, and the brown dash-dot-dotted case adopts a higher scatter at low masses. The red dashed shallow-slope model is most preferred by the data. {\em Bottom:} The effect on the $z=6$ UVLF from sharp truncations in star formation in halos of mass $M<5\times10^{9} {\rm M}_\odot$ (red dashed) and $M<2.6\times10^{10}{\rm M}_\odot$ (purple dotted, this is the mass scale below which we found a flattening in SFE is preferred) compared with our fiducial best fit model. The lensed UVLF strongly disfavors these and similar truncation models.}
  \label{fig:alt_models}
\end{figure}

The top panel of Figure \ref{fig:alt_models} shows the SFEs of our best fit enhancement models. Additionally, note that Figure \ref{fig:LF} shows the UVLF in the best fit flattening model as an illustrative case. By construction, the flattening model has a larger SFE in small halo masses and this leads to a steeper faint-end slope in the UVLF. Specifically, the $z=6$ SFE flattens at $\sim4\%$, which is a factor of several larger than the fiducial best fit SFE in the lowest masses probed by B22. Although the flattening and shallow-slope models boost the rate of ionizing photon production in small mass halos relative to our fiducial model, the changes in the reionization history are relatively small and can be compensated by lowering the uncertain escape fraction in these alternative scenarios (Figures \ref{fig:xi},\ref{fig:tau}). However, the UVLF measurements provide a less complete census of the sources of reionization (Figure \ref{fig:n_ion}) and as emphasized previously, the onset of WF coupling also occurs earlier in these models (Figure \ref{fig:rho_UV}) ($z\sim15$ in the shallow-slope case). Although not shown in the figures, the corresponding behavior in the shallow-slope case is always intermediate between those in the fiducial and flattening models.
\subsubsection{Increased Stochasticity}

In our first model with enhanced scatter in small mass halos,
we allow the variance parameter, $\sigma$, in $\phi_c(L|M,z)$ (Equation \ref{eq:CLF}) to take on a new value below $M_2$. In this case, our fitting  
prefers a value of $\sigma=1.51$ (from a prior range of $[0.05, 2.5]$) below $M_2=5\times10^9 {\rm M}_\odot$ compared to $\sigma=0.368$ in the baseline model. So the data moderately prefer enhanced scatter, although only relatively near the smallest halo mass scales probed. In this model, the average SFE is also increased. Specifically, Equations \ref{eq:f_star_scaling}-\ref{eq:f_star_normalization} show that $f_\star(M)\propto e^{\sigma^2/2}$ and so the SFE is larger in the best fit enhanced scatter model by nearly a factor of $\sim 3$ below $M_2$, as compared to the fiducial case, see the top panel of Figure \ref{fig:alt_models}. 

Next, we adopt a simplified model for a constant, non-unity duty cycle in small mass halos motivated in part by \cite{Mirocha2020,Furlanetto22_bursty}. If $0\leq f_{\rm duty}\leq1$ represents the fraction of time galaxies in those halos are actively forming stars, in the CLF this corresponds to $\phi(L|M,z)\rightarrow f_{\rm duty}\phi(L|M,z)$ for $M_{\rm min}\leq M\leq M_2$. In this model the best fit is $f_{\rm duty}=0.48$ below $M_2=10^{11} {\rm M}_\odot$. As shown in Table \ref{table:params}, some of the other fitting parameters have moderate shifts in this case -- see the top panel of Figure \ref{fig:alt_models} for the impact of this on $f_\star(M)$. As in the previous enhanced scatter model, the average SFE increases somewhat in small halos and the bursty model is likewise preferred relative to our fiducial model.

In reality, a more complicated treatment for modeling the faint end SFE may be required. A combination of the previous extensions, a smoother transition between the efficiency in small and intermediate mass halos, and additional evolution (with mass/redshift) in $\sigma$ or $f_{\rm duty}$ may be warranted. 
Ultimately, we find many promising initial candidates for extensions to the CLF model, especially the case of a shallower power-law in SFE for the lowest mass halos, which gives $\chi^2_\nu=1.32$, and the smallest AIC value (see Table \ref{table:params}). However, as discussed further in \S \ref{sec:additional} these hints are subject to systematic concerns in the lensed UVLF measurements and so we defer further exploration to future work. 

\subsection{Truncation of Star Formation at Low Mass}
\label{sec:trunc}

The UVLF measurements can also be used to strongly rule out scenarios in which the SFE drops sharply in halos below some critical mass scale, $M_2$. As we have discussed, feedback effects from supernovae and photoionization heating may strongly reduce the SFE in low mass halos. In our fiducial model, feedback leads to a gradual reduction in the SFE at small masses, while we adopt $M_{\rm min} = 10^8 {\rm M}_\odot$ as the scale below which it is strongly suppressed, as this is close to the atomic cooling mass. However, the SFE may actually be strongly reduced on mass scales somewhat larger than the atomic cooling mass: for example, photo-heating from reionization may prevent gas from accreting onto halo masses somewhat larger than our $M_{\rm min}$ value (see e.g. \citealt{Loeb2013} and references therein). B22 compares their UVLF measurements with various feedback models in the literature to constrain such cases. Here, rather than comparing with particular feedback scenarios, we simply bound the possibility that the SFE is truncated below $M_2$ and vary this scale as a free parameter. A related point is that in alternatives to CDM such as warm dark matter and fuzzy dark matter, the halo mass function itself may be suppressed at small masses (e.g. \citealt{Schive:2015kza}.) The bounds on these scenarios from the new lensed UVLF measurements will be considered in future work. 

In the simple truncation scenario considered here we set
$L_c(M,z) \rightarrow 0$ below $M_2$. Upon varying $M_2$ in our truncation model, we 
find that it is preferred to lie below the mass range probed by the measurements.\footnote{$\log_{10}(M_2/{\rm M}_\odot) =8.68\pm_{0.46}^{0.46}$ are the best fit and $1-\sigma$ errors (Table \ref{table:params}). However, the likelihood is nearly flat for $M_2\lesssim10^9{\rm M}_\odot$ and so these values alone are not as useful for determining the constraints on a truncation in the SFE.} For example, a truncation at $M_2 \gtrsim 2 \times 10^9 {\rm M}_\odot$ is excluded at $2-\sigma$ and $M_2 \gtrsim 5\times10^9{\rm M}_\odot$ is excluded at $5-\sigma$.
The bottom panel of
Figure \ref{fig:alt_models} shows example truncation models with strong suppression in halos either below $M_2=5\times10^{9}$M$_\odot$ or below the preferred $M_2$ in the flattening scenario, $M_2=2.6\times10^{10}$M$_\odot$. The other parameters have been allowed to vary in the these cases.  
The dramatic effect on the UVLF illustrates why a truncation is excluded in the mass range probed by the B21+22 data.

Additionally, the shallow-slope model of \S \ref{sec:enhance} allowed for the possibility of a steepening of $f_\star(M)$ below $M_2$ which could have indicated a more mild suppression in star formation. Instead, a shallower relation, indicating an SFE enhancement, is preferred.

These findings are similar to those in B22, which compared various UVLF models in the literature that suppress the faint end of the luminosity function. Here, we determine bounds on the mass scale of any suppression which should be agnostic to the details of particular models. Based on the B22 UVLF measurements, viable feedback models must not strongly suppress the SFE above $M \gtrsim 2 \times 10^9 {\rm M}_\odot$ (at $2-\sigma$ confidence).

\subsection{Impact of Systematic Errors on Lensed UVLF Measurements}
\label{sec:additional} 

As emphasized in \S \ref{sec:uvlf_data}, B22 recommend an additional 20\% (22\% at $z=5$) relative normalization error between the lensed and field UVLF data. This is intended to account for both systematic errors in the lensed measurements as well as cosmic variance in the relatively small fields probed behind foreground lensing clusters. 
This additional contribution to the error budget has been ignored in our analysis thus far, in part because it is non-trivial to include. In particular, while cosmic variance contributions should be essentially uncorrelated across different redshift bins, systematic errors may be correlated. Hence, a fully rigorous treatment requires knowledge of the covariance between
the uncertainties in different redshift bins. 
We can nevertheless estimate the impact of this
additional normalization error by re-running our analysis on two modified data sets, denoted here by B21+22-U and B21+22-D, where the lensed B22 measurements are all shifted up or down by 20\% (22\% at $z=5$) respectively, keeping the error bars fixed.\footnote{Some lower bounds are trimmed to forbid $\phi(m_{\rm UV})<0$ at $1-\sigma$. For measurements that are only $1-\sigma$ upper limits the upper limit is shifted instead.} These tests may overestimate the impact of this added normalization uncertainty if the offsets in fact vary with redshift (or luminosity). 

First, we adopt our fiducial CLF parameterization and fit to the shifted B21+22-U and B21+22-D data sets. We find that the baseline model performs {\em better} at explaining B21+22-D ($\chi^2_\nu=1.47$) but somewhat worse at matching B21+22-U ($\chi^2_\nu=2.04$), compared to the unmodified case ($\chi^2_\nu=1.67$). In these new fits all of the best-fit parameters vary only slightly except $p$ which controls the faint-end behavior($p=1.60(1.83)$ in the up-(down-)shifted cases). These modifications lead to only minor changes in our conclusions regarding reionization, shifting 
the timing of reionization by around $\Delta z = \pm 0.2$ at fixed escape fraction, and the onset redshift of WF coupling
by $\Delta z = \pm 0.9$.

Next we turn to the effects the added normalization uncertainties have on our alternative faint-end models. If the lensed measurements are $\sim20\%$ overestimates of their true values, the evidence for an enhancement in star formation at low-mass is much less strong. For instance, although the flattening and shallow-slope models still perform well at explaining the data ($\chi^2_\nu=1.43$ and $\chi^2_\nu=1.25$ respectively), they are less strongly preferred over the fiducial model fit to B21+22-D. For example, the shallow-slope case is still preferred (now with $s=0.35$), but the AIC difference drops from 24.3 to 15.7.
Yet, even accounting for potential systematic uncertainties in the lensed UVLF, the data seem to prefer {\em some} degree of SFE enhancement in low-mass halos relative to our fiducial case. 

Alternatively, if the lensed measurements are $\sim20\%$ {\em under-estimates} of their true value this would yield stronger evidence of enhanced SFE at low mass. For instance, a shallowing slope of $s=0.19$ would be preferred over the fiducial best fit at $\Delta$AIC$=30.7$ with respect to B21+22-U.

These shifts do not impact the conclusions regarding the truncation models. In all cases, the data disfavor strong suppressions in the SFE at low halo mass. Quantitatively, even the down-shifted data set disfavors strong suppression for halos of mass $M\geq4\times10^9 {\rm M}_\odot$ at $2-\sigma$ confidence.

In summary, even accounting for this additional systematic uncertainty, our analysis indicates that a simple power law extrapolation from $f_\star$ at $M\sim10^{10} - 10^{12}{\rm M}_\odot$ may underestimate the SFE in $M\lesssim10^{10}{\rm M}_\odot$ halos.
The exact form of the SFE in such halos is not, however, well-constrained by the current data. Further progress will likely require refinements in the lensed UVLF measurements and larger survey volumes. 

\section{Conclusion}
\label{sec:conclusion}

The aim of this work has been to fit semi-empirical models to updated analyses of the UVLF in the HFF from B22, combined with measurements in the field from B21, together spanning $z = 5-10$.
We explored the implications for our understanding of the SFE and its trends with halo mass and redshift, and the resulting consequences for our understanding of cosmic reionization, the redshifted 21 cm signal, and early JWST results.  

Our main findings can be summarized as follows. First, a simple extrapolation of the best fit results to the B21 data continues to provide a reasonably good match to the lensed UVLF measurements from B22, even though these probe up to five magnitudes fainter in UV luminosity.
The data, however, show a mild preference for an enhancement in the SFE at small masses relative to this simple extrapolation. 
Alternatively, the data may prefer an enhanced scatter in the correlation between UV luminosity and halo mass or a non-unity duty cycle at small mass scales. The combination of an enhanced average SFE and increased scatter at low masses is also plausible. However, the precise modification of the SFE in small mass halos is not yet well constrained after accounting for systematic and cosmic variance uncertainties in the lensed UVLF data. 

Next, the lensed UVLF measurements probe the SFE in halos down to $M \sim 2 \times 10^9 {\rm M}_\odot$, yet show no indication of a strong suppression on these mass scales from feedback effects. This constrains previously viable feedback models (see also B22). Further, the lensed UVLF measurements should also constrain departures from CDM in which the halo mass function is suppressed on small mass scales (e.g \citealt{Schive:2015kza}). This will be the subject of upcoming work. 

We also consider the implications of our UVLF-calibrated models for the EoR and Cosmic Dawn. Along the lines of previous work (e.g. \citealt{Robertson:2015uda,Sun2016}), we find that we can reproduce Planck 18 $\tau_e$ and current neutral fraction measurements provided the escape fraction of ionizing photons is $f_{\rm esc} = 0.1-0.2$. This scenario also roughly matches inferences of the ionizing emissivity from the Ly-$\alpha$ forest
at $z \sim 5-6$ \citep{Becker21}, although the latest results from \cite{Gaikwad:2023ubo} prefer a slightly lower escape fraction of $f_{\rm esc} = 0.05-0.1$. The consistency between our models and reionization history measurements suggests that HST measurements have, in fact, identified the sources primarily responsible for reionization. Quantitatively, even in a model with a flat SFE in small mass halos the lensed UVLF measurements account for $\gtrsim 50\%$ of the total ionizing photon budget at $z \leq 9$, with some caveats discussed in \S \ref{sec:census}. In our fiducial scenario, WF coupling (related to the onset of the 21 cm absorption signal) occurs near $z \sim 13$, while our flat SFE model gives an onset redshift ($z \sim 18$) similar to that suggested by the possible EDGES 21 cm absorption signal \citep{Bowman:2018yin}; see also \cite{Mirocha:2018cih}.

Finally, our HST-calibrated models agree well with early JWST results at $z \leq 10$, but current UVLF estimates from JWST lie quite a bit above the HST-calibrated models at higher redshift. 
This might be reconciled if the SFE reverses the decreasing trend with redshift suggested by the HST data, and increases at the higher redshifts only probed by JWST ($z \geq 10$). However, dramatic evolution would be required. We discuss alternative resolutions further in \S \ref{sec:jwst}. 

In the near future, upcoming JWST lensed UVLF measurements should extend the reach of the lensed estimates beyond $z \geq 10$. This will provide further valuable information regarding the SFE at high redshifts and low halo masses, while early JWST candidates at $z \geq 10$ have thus far been confined to the bright-end of the UVLF. In addition, we look forward to spectroscopic confirmations of photometric candidates from JWST, and further assessments of the purity of Lyman-break selected samples at $z \gtrsim 10$. The UVLF estimates can also be combined with Balmer line measurements from JWST, which provide a more direct handle on the ionizing emissivity, while upcoming line-intensity mapping surveys may probe the collective impact of individually faint galaxies \citep{Bernal:2022jap}. Further, we anticipate improved measurements of the reionization history itself from a wide variety of observational probes. Finally, additional comparisons between semi-empirical models and first-principles hydro-dynamical simulations of galaxy formation will also improve our understanding of the sources of reionization. 

\section*{Acknowledgements}
We thank Jason Sun for helpful discussions. We also thank the anonymous referee for helpful suggestions which improved the presentation of our results. We acknowledge support through NASA ATP grant 80NSSC20K0497 and the Kaufman foundation. 
\software{\texttt{Astropy} \citep{astropy2013, astropy2018, astropy2022}, \texttt{NumPy} \citep{numpy}, \texttt{SciPy} \citep{2020SciPy-NMeth}, \texttt{emcee} \citep{Foreman-Mackey12}, \texttt{CAMB} \citep{Lewis_2000}, \texttt{Matplotlib} \citep{matplotlib}, \texttt{corner} \citep{Foreman-Mackey2016}}

\section*{Data availability}
The data underlying this article is available upon request. 

\bibliography{mybib}
\bibliographystyle{aasjournal}

\end{document}